\documentclass[10pt,journal,twocolumns]{IEEEtran}
\usepackage{subfigure}
\usepackage{amssymb}
\usepackage{amsthm}
\usepackage{amsmath}
\usepackage{hyperref}
\usepackage{graphicx}
\usepackage{colortbl,dcolumn}
\usepackage{booktabs}
\usepackage{xcolor}
\usepackage{cite}
\usepackage{threeparttable}

\usepackage{algorithm}
\usepackage{algorithmic}

\newtheorem{definition}{Definition}
\newtheorem{lemma}{Lemma}
\newtheorem{theorem}{Theorem}
\newtheorem{proposition}{Proposition}
\newtheorem{corollary}{Corollary}
\newtheorem{remark}{Remark}
\newtheorem{problem}{Problem}

\newcommand{\defref}[1]{Definition~\ref{#1}}
\newcommand{\lemref}[1]{Lemma~\ref{#1}}
\newcommand{\thmref}[1]{Theorem~\ref{#1}}
\newcommand{\propref}[1]{Proposition~\ref{#1}}
\newcommand{\corref}[1]{Corollary~\ref{#1}}
\newcommand{\figref}[1]{Fig.~\ref{#1}}
\newcommand{\tabref}[1]{Table~\ref{#1}}
\newcommand{\secref}[1]{Section~\ref{#1}}
\newcommand{\apxref}[1]{Appendix~\ref{#1}}
\newcommand{\probref}[1]{Problem~\ref{#1}}

\newcommand{\algref}[1]{Algorithm~\ref{#1}}
\newcommand{\lineref}[1]{Line~\ref{#1}}

\usepackage{mathtools} 

\usepackage{enumerate}

\ifCLASSOPTIONcompsoc
\else
\fi
\ifCLASSINFOpdf
\else
\fi

\graphicspath{{pdfs/},{eps/},{authorphotos/}}

\begin{document}
%
\title{Finite-Horizon Throughput Region for Wireless Multi-User Interference Channels}
%
%
%
%

\author{Yirui~Cong,~\IEEEmembership{Student~Member,~IEEE,}
        Xiangyun~Zhou,~\IEEEmembership{Member,~IEEE,}
        and~Rodney~A.~Kennedy,~\IEEEmembership{Fellow,~IEEE}
\IEEEcompsocitemizethanks{\IEEEcompsocthanksitem

Y. Cong, X. Zhou and R. Kennedy are with the Research School
of Engineering, Australian National University, Australia (Email: \{yirui.cong, xiangyun.zhou, Rodney.Kennedy\}@anu.edu.au).\protect
}
}

\IEEEtitleabstractindextext{%
\begin{abstract}

This paper studies a wireless network consisting of multiple transmitter-receiver pairs where interference is treated as noise. Previously, the throughput region of such networks was characterized for either one time slot or an infinite time horizon. We aim to fill the gap by investigating the throughput region for transmissions over a finite time horizon. Unlike the infinite-horizon throughput region, which is simply the convex hull of the throughput region of one time slot, the finite-horizon throughput region is generally non-convex. Instead of directly characterizing all achievable rate-tuples in the finite-horizon throughput region, we propose a metric termed the rate margin, which not only determines whether any given rate-tuple is within the throughput region (i.e., achievable or unachievable), but also tells the amount of scaling that can be done to the given achievable (unachievable) rate-tuple such that the resulting rate-tuple is still within (brought back into) the throughput region. Furthermore, we derive an efficient algorithm to find the rate-achieving policy for any given rate-tuple in the finite-horizon throughput region.

\end{abstract}

}

\maketitle

\IEEEdisplaynontitleabstractindextext

%
\IEEEpeerreviewmaketitle

\section{Introduction}


\subsection{Motivation}\label{sec:Motivation}

The capacity region of a general wireless multi-user network is largely an open research problem.
Since the information-theoretic capacity of a multi-user wireless network is extremely challenging to study due to many unsolved problems of network information theory, an alternative approach has been taken from a network-layer perspective to analyze the set of all achievable rates between the communication pairs of the network under any given modulation and coding strategy~\cite{NeelyM2005JSAC}.
Such studies commonly assume that the interference in the network is treated as noise, hence the capacity of each link is determined by signal-to-interference-plus-noise ratio (SINR).
Even under the assumption of treating interference as noise, the set of all achievable rate-tuples in a multi-user network, which we will name as throughput region\footnote{In~\cite{NeelyM2005JSAC} and other related work, the throughput region is also called the (network-layer) capacity region. The reason for using the nomenclature ``throughput region'' is to distinguish it from the capacity region in the information theoretic sense.} in this work, is still not well understood.

Interference from concurrent transmissions leads to highly nonlinear couplings among transmitter-receiver pairs, which makes the throughput region difficult to determined.
Many studies have been devoted to maximizing the sum rate or proportional fairness, with either centralized or distributed power control algorithms, and typically consider one time slot\footnote{A time slot is the duration of a codeword consisting of multiple channel uses.} only (see~\cite{GesbertD2007,TanC2013} and references therein).
Apart from interference, another key challenge is from the consideration of multiple time slots.
It is well known that, in a point-to-point system, knowing the achievable rate in one time slot is sufficient to derive the achievable rate for any number of time slots. However, this is not the case for networks with multiple transmitter-receiver pairs where the couplings of their transmission policies among multiple time slots must be taken into account.
Indeed, the multi-slot throughput region is generally larger than the single-slot throughput region~\cite{GeorgiadisL2006BOOK} for multi-user interference channels.
Noticing this challenge, the set of all achievable rate-tuples in multi-user wireless networks was studied for the case of an infinite number of time slots~\cite{TassiulasL1992,TassiulasL1991Thesis}, which we name as the infinite-horizon throughput region\footnote{In this paper, the term ``infinite horizon'' refers to an infinite number of time slots and ``finite horizon'' refers to a finite number of time slots.}.

Despite the significant efforts made on studying the achievable rate-tuple and throughput region for both one time slot and infinite horizon, significantly less is known about the throughput region over a finite horizon.
In wireless networks, the network traffic, channel condition and even network topology change with time~\cite{GeorgiadisL2006BOOK}.
It is desirable to design transmission for a finite time duration such that the network and channel information used in the design is timely and matches with the condition during the actual transmission.
Moreover, there are many wireless applications in which the nodes only communicate for a short period of time, e.g., wireless sensor networks~\cite{YickJ2008} where sensors have a short period of transmission mode followed by sleep mode.
Therefore, it is necessary to study the finite-horizon throughput region which directly tells whether a given rate-tuple can be achieved or not within any given number of time slots.
To the best of our knowledge, the finite-horizon throughput region of a multi-user wireless network has not yet been investigated.

Another important reason for considering the finite-horizon throughput region is the guaranteed delay.
For example, if a rate-tuple is achievable in a five-slot throughput region, then the time delay for the transmitted packets is at most five time slots.
On the contrary, any rate-tuple in an infinite-horizon throughput region can possibly cause an unacceptably large delay. This also motivates us to study the finite-horizon throughput region of a multi-user wireless network.

\subsection{Related Work}

Since this is the first work that rigorously studies the finite-horizon throughput region, the most related prior works are the ones on infinite-horizon throughput region. Specifically, the seminal work in~\cite{TassiulasL1992,TassiulasL1991Thesis} introduced the infinite-horizon throughput region and gave two important results: the infinite-horizon throughput region is the convex hull of one-slot throughput region; and the max-weight algorithm can achieve any given rate-tuple in the throughput region.
In~\cite{NeelyM2005JSAC}, the infinite-horizon throughput region was generalized and applied to time-varying wireless networks, and a max-weight algorithm based transmission policy was designed.
We recommend the tutorials in~\cite{GeorgiadisL2006BOOK,LinX2006,NeelyM2010} to readers who are interested in the infinite-horizon throughput region.

Some recent studies focused on reducing the delay by shrinking the infinite-horizon throughput region~\cite{KarK2012,BoyaciC2013,NeelyM2013,XueD2013}, where the average delay was studied in~\cite{KarK2012,BoyaciC2013}, and the worst-case delay was analyzed in~\cite{NeelyM2013,XueD2013}.
It was observed by~\cite{KarK2012,BoyaciC2013,NeelyM2013,XueD2013} that choosing a rate-tuple closer to the boundary of the infinite-horizon throughput region causes a larger delay, and hence, shrinking the throughput region removes those rate-tuples near the boundary corresponding to large delays.
Furthermore, the effect of finite buffer size was consider in~\cite{LeL2012,XueD2013VT,XueD2015}, and it turned out that the required buffer size increases with the rate-tuple.
Indeed, by Little's law, the average length of data queue is proportional to the delay. Hence, this line of work also demonstrated the delay caused by the rate-tuples in the infinite-horizon throughput region.

We stress that in light of the studies on infinite-horizon throughput region, a small number of studies have introduced the concept of finite-horizon throughput region.
However, they did not analyze any property of the finite-horizon throughput region:
The work in~\cite{NeelyM2011INFOCOM} proposed a $T$-slot lookahead utility which helped to analyze the short-term performance for the proposed opportunistic scheduling algorithm, but no analysis on finite-horizon throughput region was presented;
In~\cite{NeelyM2013}, the rate-tuple over a finite time horizon was defined, but it was only employed to derive the infinite-horizon throughput region when the number of time slots goes to infinity.
A possible reason for the lack of study on the finite-horizon throughput region might be that the finite-horizon throughput region was thought to have similar properties as its infinite-horizon counterpart. As we will discuss in this work, however, the finite-horizon throughput region behaves very differently as compared with the infinite-horizon throughput region.

\subsection{Our Contributions}

In this work, we investigate the finite-horizon throughput region of a wireless network consisting of multiple transmitter-receiver pairs.
It should be noted that studying the finite-horizon throughput region is far more challenging than its infinite-horizon counterpart for the following reasons:
(i) Unlike the convex throughput region for the infinite horizon, the finite-horizon throughput region is non-convex and the computational complexity for determining it is exponentially increasing with the number of time slots.
(ii) As we will show, a rate-tuple that is achievable in $T_1$ time slots may not be achievable in $T_2 \,\,(T_2 > T_1)$ time slots.
This is in contrast with the fact that any achievable rate-tuple over a finite horizon is also achievable over the infinite horizon.
This property prevents us from using a result for one throughput region to obtain a result for another throughput region with a different number of time slots.

Instead of directly characterizing the throughput region by finding the set of all achievable rate-tuples, we provide an efficient method to determine whether an arbitrary given rate-tuple is achievable or not. More specifically:
\begin{itemize}
\item   We propose a metric termed the \textit{rate margin}. By computing the rate margin of any given rate-tuple, we are able to tell whether a given rate-tuple is achievable within the considered finite horizon. Furthermore, the rate margin also provides information to the system designer on: (i) how much one can scale up the given achievable rate-tuple so that the resulting rate-tuple is still within the finite-horizon throughput region; (ii) how much one can scale down the given unachievable rate-tuple so that the resulting rate-tuple is brought back into the finite-horizon throughput region.

\item   We provide the rate-achieving policy for any achievable rate-tuple in a finite-horizon throughput region by determining the transmit power and rate for each communication pair in each time slot.

\item   We formulate an optimization problem for computing the rate margin and deriving the rate-achieving policy. The solution inevitably requires a search. To reduce the complexity while maintain the optimality of the search, we use three techniques among which the most important one is the proposed admissible heuristic function that allows the highly-efficient A* search algorithm to be employed. The simulation result demonstrates the computational efficiency of our algorithm.

\end{itemize}

\subsection{Paper Organization}

The system model and problem description are given in \secref{sec:System Model} and \secref{sec:Problem Description: Rate Margin and Rate-Achieving Policy}, respectively.
In \secref{sec:Derivation of Rate Margin and Rate-Achieving Policy}, an optimization problem is defined to solve rate margin and rate-achieving policy, and an efficient solution method for this problem is also proposed.
The numerical examples are given in \secref{sec:Simulation Examples} to illustrate the effectiveness of our approach and corroborate our analytical results.

\subsection{Notation}

Throughout this paper, for a vector $\mathbf{a} = [a^{(1)},\dotsc,a^{(N)}]^{\mathrm{T}}$ (where $\mathrm{T}$ is the transpose operator), $(\mathbf{a})^+$ denotes $\max\{a^{(n)},0\}$ for all $n\in\{1,\dotsc,N\}$.
The cardinality of a set $\mathcal{A}$ is $|\mathcal{A}|$.
For $\mathbf{x}_1 = [x_1^{(1)},\dotsc,x_1^{(N)}]^{\mathrm{T}}$ and $\mathbf{x}_2 = [x_2^{(1)},\dotsc,x_2^{(N)}]^{\mathrm{T}}$, $\mathbf{x}_1\succeq$ ($\succ,\preceq,\prec$) represents $x_1^{(n)}\geq$ ($>,\leq,<$) $x_2^{(n)}$ for all $n\in\{1,\dotsc,N\}$.
$\mathbf{x}_1 \succneqq$ ($\precneqq$) $\mathbf{x}_2$ means $\mathbf{x}_1 \succeq$ ($\preceq$) $\mathbf{x}_2$ but $\mathbf{x}_1 \neq \mathbf{x}_2$.
$\overline{\mathbb{R}}_{+}^{N}$ ($\mathbb{R}_+^N$) means $\left\{\mathbf{x} \in \mathbb{R}^N: \mathbf{x} \succeq \, (\succ) 0\right\}$.
$\mathbf{0}$ stands for the zero vector with proper dimension.
For $x \in \mathbb{R}$, $\lfloor x\rfloor$ returns the largest integer not greater than $x$.

\section{System Model and Throughput Region}\label{sec:System Model}

\begin{figure}[h]
\centering
\includegraphics [width=0.8\columnwidth]{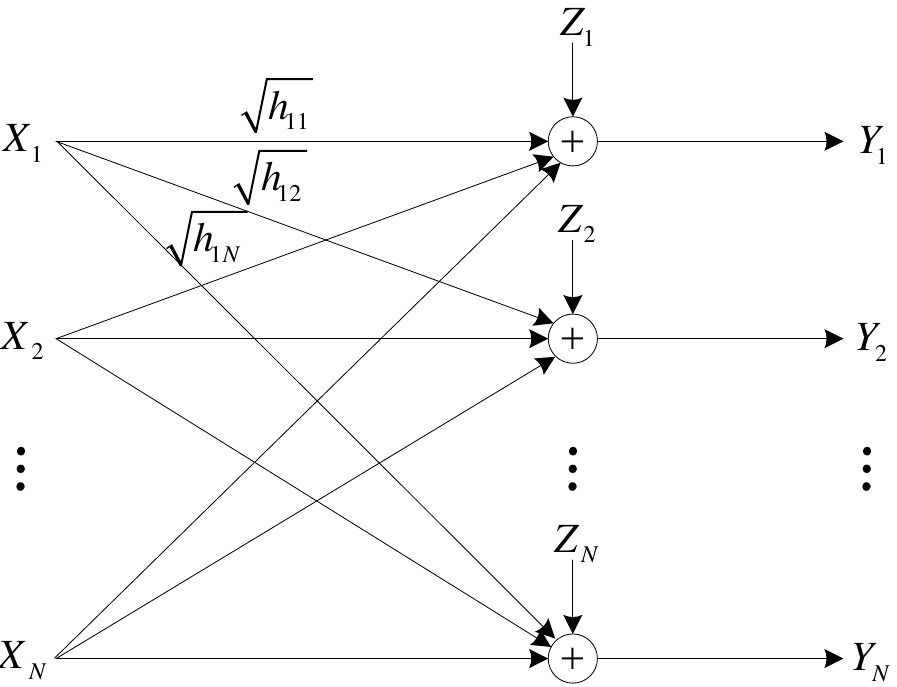}
\caption{Channel model for $n$ transmitter-receiver pairs.}
\label{fig:Multi-user Gaussian interference channel.}
\end{figure}

We consider $N$ transmitter-receiver pairs in a wireless network, where $\mathrm{Tx}_n$ and $\mathrm{Rx}_n$ denote the transmitter and receiver of the $n$\textsuperscript{th} communication pair.
The channel is memoryless, and the relationship between channel input $X_m \in \mathbb{R}$ (corresponding to $\mathrm{Tx}_m$, and $m\in\{1,\dotsc,N\}=:\mathcal{N}$) to output $Y_n$ ($n \in \mathcal{N}$) is
\begin{equation}\label{eqn:Relationship between Channel Inputs and Outputs}
Y_n = \sum_{m \in \mathcal{N}} \sqrt{h_{mn}} X_m + Z_n, \quad n \in \mathcal{N},
\end{equation}
where $h_{mn}$ is the channel gain between $\mathrm{Tx}_m$ and $\mathrm{Rx}_n$; and $Z_n$ is Gaussian white noise with power $W_n$.
When decoding, $\mathrm{Rx}_n$ treats the interference $\sum_{m \neq n} \sqrt{h_{mn}} X_m$ as noise.
We can see that our channel is indeed a multi-user Gaussian interference channel, as shown in \figref{fig:Multi-user Gaussian interference channel.}.

The time is slotted and each time slot contains $L$ channel uses for transmitting and receiving a codeword (i.e., the length of a codeword, or simply the blocklength, is $L$).
We consider a finite time horizon of $T$ time slots and assume that the channel gains $h_{mn}$ remain constant over the $T$ time slots.
In each time slot, every transmitter-receiver pair chooses to transmit or not.
That is, for time slot $t\in\{1,\dotsc,T\}=:\mathcal{T}$, the transmitter $\mathrm{Tx}_n$ ($n\in\mathcal{N}$) chooses its transmit power $s_t^{(n)}$ from the transmit-power set $\mathcal{S}^{(n)}$, where $0$ is included for representing no transmission.
%
%
%
Since the number of available transmit power options in a practical communication system is usually finite (e.g., see the discrete power control in~\cite{XingY2008,ZhangH2011}), we model $\mathcal{S}^{(n)}$ as a finite set.
Furthermore, we label $\mathbf{s}_t = \big[s_t^{(1)},\dotsc,s_t^{(N)}\big]^{\mathrm{T}}$, and $\mathcal{S}: = \mathcal{S}^{(1)}\times\dotsb\times\mathcal{S}^{(N)}$.
Hence, $\mathbf{s}_t \in \mathcal{S}$, and we call $\mathcal{S}$ the transmit-power-tuple set.

For time slot $t$, the SINR for each transmitter-receiver pair is determined by
\begin{align}\label{eqn:SINR}
\gamma_n(\mathbf{s}_t) = \frac {h_{nn} s_t^{(n)}} {W_n + \sum_{m\neq n} h_{mn} s_t^{(m)}}, \quad n,m\in\mathcal{N},
\end{align}
%
%
where the interference is treated as noise.
Given the SINR $\gamma_n(\mathbf{s}_t)$, blocklength $L$, and error probability $\epsilon$, the maximum achievable rate for transmitter-receiver pair $n$ is defined as
\begin{equation}\label{eqn:Maximal Channel Coding Rate}
\mu_{\max}^{(n)}(\gamma_n(\mathbf{s}_t), L, \epsilon) = \frac {1}{L} \log_2 M^*(L, \epsilon),
\end{equation}
where $M^*(L, \epsilon)$ represents the maximal code size as defined in~\cite{PolyanskiyY2010,TanV2014}.
In our numerical results, we use the following accurate approximation of $\mu_{\max}^{(n)}(\gamma_n(\mathbf{s}_t), L, \epsilon)$ for Gaussian channels~\cite{PolyanskiyY2010}:
\begin{equation}\label{eqn:Approximation of Maximum Rate under Finite Blocklength Effect}
\mu_{\max}^{(n)}(\gamma_n(\mathbf{s}_t), L, \epsilon) \approx \frac {1} {2} \log_2 (1 + \gamma_n(\mathbf{s}_t)) - \sqrt{\frac {V}{L}} Q^{-1}(\epsilon),
\end{equation}
where $Q^{-1}$ is the inverse of standard Gaussian complimentary CDF, and $V$ is the channel dispersion
\begin{equation}\label{eqn:Channel Dispersion}
V = \frac {\log_2^2 e}{2} \left[1 - \frac {1}{(1 + \gamma_n(\mathbf{s}_t))^2}\right].
\end{equation}
We stress that all the analytical results in this paper hold for both the generic expression in~\eqref{eqn:Maximal Channel Coding Rate} and the explicit approximation in~\eqref{eqn:Approximation of Maximum Rate under Finite Blocklength Effect}, where \eqref{eqn:Approximation of Maximum Rate under Finite Blocklength Effect} is primarily used to obtain numerical results.

For transmitter-receiver pair $n$, in time slot $t$, any rate $\mu_t^{(n)} \in \left[0,~\mu_{\max}^{(n)}(\gamma_n(\mathbf{s}_t), L, \epsilon)\right]$ is achievable, i.e., there exist some channel codes with rate $\mu_t^{(n)}$ such that the decoding error probability is bounded by $\epsilon$.
Under a given transmit-power-tuple $\mathbf{s}_t$, all achievable rate-tuples for $N$ transmitter-receiver pairs form the following set
\begin{multline}\label{eqn:All Achievable Rates Given a Transmit-Power-Vector}
\Big\{\left[\mu_t^{(1)},\ldots,\mu_t^{(N)}\right]\colon\\
\mu_t^{(n)} \in \left[0,~\mu_{\max}^{(n)}(\gamma_n(\mathbf{s}_t), L, \epsilon)\right], \quad n \in \mathcal{N}\Big\}.
\end{multline}
By defining ${\boldsymbol \mu}_t = \left[\mu_t^{(1)},\ldots,\mu_t^{(N)}\right]$ and
\begin{multline}\label{eqn:Maximum rate-tuple for a given transmit-power-tuple}
{\boldsymbol \mu}_{\max}(\mathbf{s}_t, L, \epsilon) \\= \left[\mu_{\max}^{(1)}(\gamma_1(\mathbf{s}_t), L, \epsilon),\ldots,\mu_{\max}^{(N)}(\gamma_N(\mathbf{s}_t), L, \epsilon)\right],
\end{multline}
equation \eqref{eqn:All Achievable Rates Given a Transmit-Power-Vector} can be rewritten in the compact form
\begin{equation}\label{eqn:All Achievable Rates Given a Transmit-Power-Vector Rewritten}
\left\{{\boldsymbol \mu}_t\colon {\boldsymbol \mu}_t \preceq {\boldsymbol \mu}_{\max}(\mathbf{s}_t, L, \epsilon)\right\}.
\end{equation}
Note that \eqref{eqn:All Achievable Rates Given a Transmit-Power-Vector Rewritten} contains all achievable rate-tuples in one time slot for one transmit-power-tuple $\mathbf{s}_t$.
Then, the $1$-slot throughput region is defined as the region of all achievable rate-tuples for all possible transmit-power-tuples, and the $1$-slot throughput region for time slot $t$ is
\begin{align}\label{eqn:One-Slot Throughput Region}
\Lambda_{[1],t} = \bigcup_{\mathbf{s}_t \in \mathcal{S}}\left\{{\boldsymbol \mu}_t\colon 0 \preceq {\boldsymbol \mu}_t \preceq {\boldsymbol \mu}_{\max}(\mathbf{s}_t, L, \epsilon)\right\},
\end{align}
where $\mathcal{S}$ is the set of all possible transmit-power-tuples.
Note that $\Lambda_{[1],t}$ are the same for all $t$, and thus, for simplicity, we label $\Lambda_{[1],1} = \dotsb = \Lambda_{[1],T} = \Lambda_{[1]}$.

Similar to the one-slot throughput region, the finite-horizon throughput region for $T$ time slots is defined as follows.

\begin{definition}[Finite-Horizon Throughput Region]\label{def:Finite-Slots Throughput Region}
The $T$-slot throughput region $\Lambda_{[T]}$ is the set of average rate-tuples that can be achieved in $T$ time slots, i.e.,
\begin{align}\label{eqn:Finite-Horizon Throughput Region}
\Lambda_{[T]} = \left\{{\boldsymbol \mu}_{[T]}\colon {\boldsymbol \mu}_{[T]} = \frac{1}{T}\sum_{t=1}^{T} {\boldsymbol \mu}_t,\quad{\boldsymbol \mu}_t \in \Lambda_{[1]}\right\}.
\end{align}
\end{definition}

For a clear illustration of finite-horizon throughput region, we define the weak Pareto frontier and Pareto frontier in \defref{def:Weak Pareto Frontier and Pareto Frontier} which also plays an important role in the analytical results in this paper.

\begin{definition}[Weak Pareto Frontier and Pareto Frontier]\label{def:Weak Pareto Frontier and Pareto Frontier}
For a given set $\mathcal{A}$, the weak Pareto frontier is
\begin{align}\label{eqn:Weak Pareto Frontier}
\mathcal{B} = \left\{b\in\mathcal{A}\colon\left\{a\in\mathcal{A}\colon a \succ b \right\} = \emptyset\right\},
\end{align}
and the Pareto Frontier is
\begin{align}\label{eqn:Pareto Frontier}
\overline{\mathcal{B}} = \left\{b\in\mathcal{A}\colon\left\{a\in\mathcal{A}\colon a \succeq b \right\} = \{b\}\right\}.
\end{align}
It should be noted that $\overline{\mathcal{B}} \subseteq \mathcal{B}$.
\end{definition}

With \defref{def:Weak Pareto Frontier and Pareto Frontier}, we define the weak Pareto frontier and Pareto frontier of $\Lambda_{[T]}$ as $\mathcal{M}_{[T]}$ and $\overline{\mathcal{M}}_{[T]}$, respectively.
Additionally, we say the rate-tuples on the weak Pareto frontier are the boundary rate-tuples.

\begin{figure*}[htb]
\centering
\subfigure[]{\includegraphics [width=0.6\columnwidth]{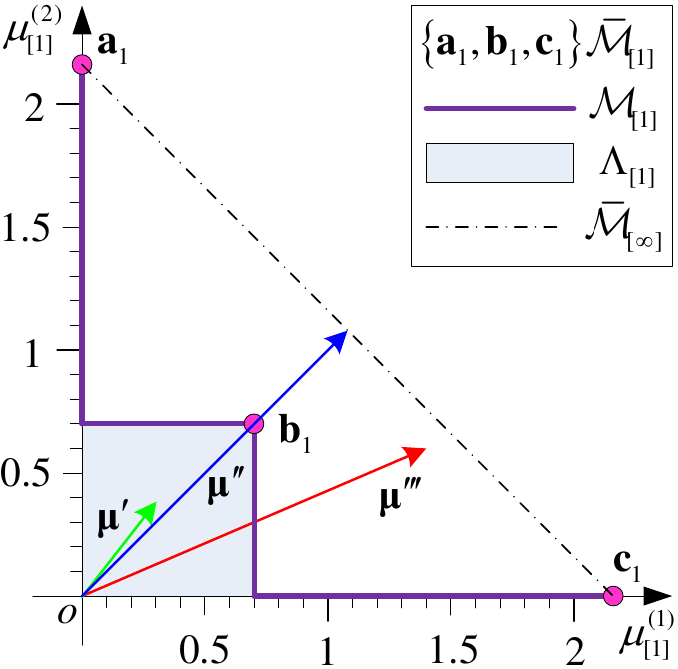}\label{fig:Example for Throughput Region 1}}\hspace{0.1em}
\subfigure[]{\includegraphics [width=0.6\columnwidth]{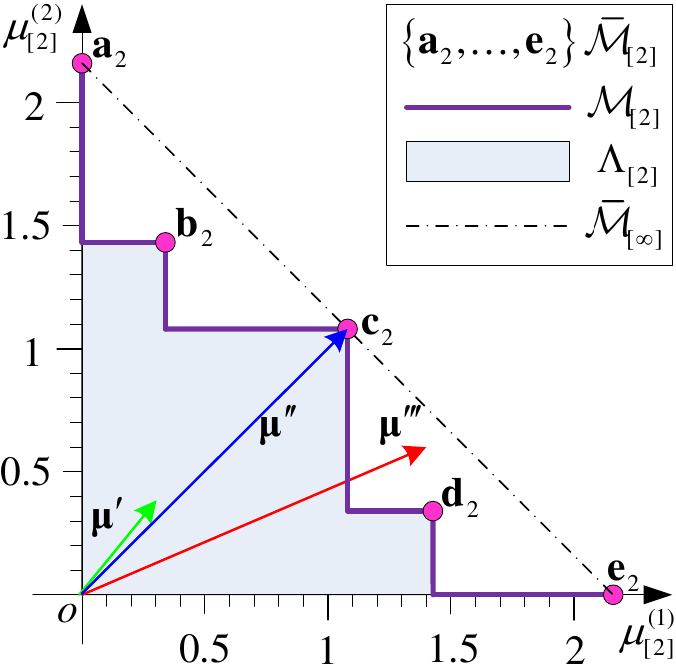}\label{fig:Example for Throughput Region 2}}\hspace{0.1em}
\subfigure[]{\includegraphics [width=0.6\columnwidth]{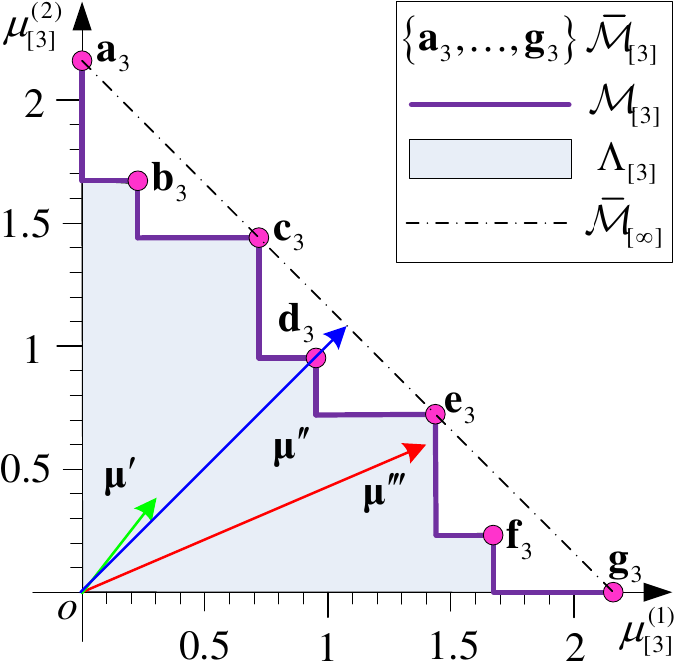}\label{fig:Example for Throughput Region 3}}\hspace{0.1em}

\caption{Examples of throughput regions of two transmitter-receiver pairs.
The network parameters are: channel gains $h_{11} = h_{22} = 1$, $h_{12} = h_{21} = 0.3$, transmit-power sets $\mathcal{S}^{(1)} = \mathcal{S}^{(2)} = \{0, 3\}$, powers of white noises $W_1 = W_2 = 0.1$, blocklength $L = 100$, and error probability $\epsilon = 0.001$.
By using the maximum rate approximation in~(4), the $1$, $2$, and $3$-slot throughput regions are shown in (a), (b), and (c), respectively.
We consider rate-tuples ${\boldsymbol \mu}' = [0.3,0.4]^{\mathrm{T}}$, ${\boldsymbol \mu}'' = [1.08,1.08]^{\mathrm{T}}$, and ${\boldsymbol \mu}''' = [1.4,0.6]^{\mathrm{T}}$.
In each finite-horizon throughput region, the pink circles compose the Pareto frontier, and the purple (thick) lines form the weak Pareto frontier, and the shaded area is the interior of a throughput region.
Note that the finite-horizon throughput region includes both the interior and weak Pareto frontier (Pareto frontier is in the weak Pareto frontier as mentioned in Definition~2).
For the dash-dotted lines, they are the Pareto frontier of the well-known infinite-horizon throughput region (see~[4]), which is the convex hull of $\Lambda_{[1]}$.
}\label{fig:Examples for throughput region}
\end{figure*}

Three examples are given in \figref{fig:Examples for throughput region} to illustrate the shape of finite-horizon throughput region.
We consider two transmitter-receiver pairs, so the throughput regions are in two dimensions.
The detailed network parameters are given in the caption of \figref{fig:Examples for throughput region}.
Using the same network parameters, \figref{fig:Example for Throughput Region 1}, \figref{fig:Example for Throughput Region 2}, and \figref{fig:Example for Throughput Region 3} illustrate the throughput regions for $T = 1$, $T = 2$, and $T = 3$, respectively.
We use the rate-tuples ${\boldsymbol \mu}'$, ${\boldsymbol \mu}''$, and ${\boldsymbol \mu}'''$ (whose values are given in the caption of \figref{fig:Examples for throughput region}) to compare the differences among finite-horizon throughput regions for different $T$:
\begin{itemize}
\item   ${\boldsymbol \mu}'$ is in $\Lambda_{[1]}$, $\Lambda_{[2]}$ and $\Lambda_{[3]}$.
\item   ${\boldsymbol \mu}''$ is in $\Lambda_{[2]}$, but not in $\Lambda_{[1]}$ or $\Lambda_{[3]}$.
\item   ${\boldsymbol \mu}'''$ is in $\Lambda_{[3]}$, but not in $\Lambda_{[1]}$ or $\Lambda_{[2]}$.
\end{itemize}
We take ${\boldsymbol \mu}'''$ as an example.
From the caption, ${\boldsymbol \mu}''' = [1.4,0.6]^{\mathrm{T}}$ and it can be achieved within $3$ time slots by
\begin{align}\label{eqn:Example for Average Rate-Tuple}
{\boldsymbol \mu}''' = \frac {1}{3} \left(\begin{bmatrix}2.1\\0\end{bmatrix} + \begin{bmatrix}2.1\\0\end{bmatrix} + \begin{bmatrix}0\\1.8\end{bmatrix}\right),
\end{align}
which means letting communication pair $1$ transmitting at the rate of $2.1$ in the first two time slots and communication pair $2$ transmitting at the rate of $1.8$ in the third time slot.
Note that ${\boldsymbol \mu}'''$ is not achievable within $T = 1$ or $T = 2$ time slots.
Intuitively, if $T_1 < T_2$, the relationship between their throughput regions seems to be $\Lambda_{[T_1]} \subseteq \Lambda_{[T_2]}$ (e.g., $\Lambda_{[1]} \subseteq \Lambda_{[2]}$).
However, this intuition turns out to be incorrect, as ${\boldsymbol \mu}''$ is in $\Lambda_{[2]}$ but not in $\Lambda_{[3]}$.

\begin{remark}[Uncertain-Inclusion Property]\label{rek:Uncertain-Inclusion Property}
For the convenience of discussion, we label $\lim_{T \rightarrow \infty} \Lambda_{[T]}$ as $\Lambda_{[\infty]}$.
It is easy to verify that $\Lambda_{[T]} \subseteq \Lambda_{[\infty]}$ holds for any finite $T$, since $\Lambda_{[\infty]}$ is a convex hull of $\Lambda_{[1]}$ (as shown in~\cite{GeorgiadisL2006BOOK}).
However, in general, if $T_1 < T_2$, the proposition $\Lambda_{[T_1]} \subseteq \Lambda_{[T_2]}$ does not always hold true.
We call this property the uncertain-inclusion property of the finite-horizon throughput region.
Specifically, it can be proved that if $T_1$ is an factor of $T_2$, then $\Lambda_{[T_1]} \subseteq \Lambda_{[T_2]}$; but if $T_1$ is not an factor of $T_2$, then $\Lambda_{[T_1]} \subseteq \Lambda_{[T_2]}$ does not hold in general, which highly depends on the structure of $\Lambda_{[1]}$.

The uncertain-inclusion property prevents us from analyzing $\Lambda_{[T_2]}$ based on the information from $\Lambda_{[T_1]}$ in general, and therefore we cannot determine whether a rate-tuple achievable in $T_1$ slots is still achievable in $T_2$ slots.
\end{remark}

\section{Problem Description: Rate Margin and Rate-Achieving Policy}\label{sec:Problem Description: Rate Margin and Rate-Achieving Policy}

In this work, we propose an important metric to characterize $\Lambda_{[T]}$, termed the rate margin.
The rate margin has three useful properties:
\begin{itemize}
\item   The rate margin determines whether a given rate-tuple is achievable or not within $T$ time slots.
\item   If a rate-tuple is achievable, the rate margin gives the headroom for scaling up the rate-tuple that remains achievable.
\item   Similarly, if a rate-tuple is unachievable, the rate margin tells exactly by what extent the rate-tuple needs to be scaled down to be achievable.
\end{itemize}
We also study the rate-achieving policy, which gives a method to achieve any given rate-tuple in $\Lambda_{[T]}$.

Now, we give the definition of the rate margin.

\begin{definition}[Rate Margin]\label{def:Rate Margin}
For a $1$-slot throughput region $\Lambda_{[1]}$ and $T$ time slots, the rate margin $\delta_T(\cdot)\colon \overline{\mathbb{R}}_+^N \rightarrow \mathbb{R}_+\bigcup\{\infty\}$ is a function of rate-tuple ${\boldsymbol \mu}_{[T]}$ that
\begin{align}\label{eqn:Rate Margin}
\delta_T({\boldsymbol \mu}_{[T]}) = \underset{{\boldsymbol \mu}'_{[T]} \in \overline{\mathcal{M}}_{[T]}~~}{\max} \underset{n\in\mathcal{N}} {\min} \left\{\frac{\mu'^{(n)}_{[T]}}{\mu^{(n)}_{[T]}}\right\},
\end{align}
where $\mu^{(n)}_{[T]}$ and $\mu'^{(n)}_{[T]}$ are the $n$\textsuperscript{th} component of ${\boldsymbol \mu}_{[T]}$ and ${\boldsymbol \mu}'_{[T]}$, respectively.
Note that the rate margin can be infinite.
\end{definition}

The rate margin has several useful properties, which are given in \propref{prop:Completeness of Rate Margin}, \propref{prop:Achievable Rate's Maximum Magnification for Remaining Achievable}, \propref{prop:Unachievable Rate's Minimum Minification for Becoming Achievable}, and \corref{cor:Boundary Rate Completeness of Rate Margin}.
Firstly, the rate margin can be used to determine whether a given rate-tuple is achievable or not:

\begin{proposition}\label{prop:Completeness of Rate Margin}
$\forall {\boldsymbol \mu}_{[T]} \in \Lambda_{[T]}$ if and only if $\delta_T({\boldsymbol \mu}_{[T]}) \geq 1$.
\end{proposition}

\begin{IEEEproof}
Necessity.~$\forall {\boldsymbol \mu}_{[T]} \in \Lambda_{[T]}$, there exists at least one ${\boldsymbol \mu}'_{[T]} \in \overline{\mathcal{M}}_{[T]}$ such that
\begin{align}
\underset{n\in\mathcal{N}} {\min} \left\{\frac{\mu_{[T]}'^{(n)}}{\mu^{(n)}_{[T]}}\right\} \geq 1,
\end{align}
since it otherwise contradicts~\eqref{eqn:Pareto Frontier} in the Pareto frontier definition.
Thus, $\delta_T({\boldsymbol \mu}_{[T]}) \geq 1$ in~\eqref{eqn:Rate Margin}.

Sufficiency.~If $\delta_T({\boldsymbol \mu}_{[T]}) \geq 1$, but we assume ${\boldsymbol \mu}_{[T]} \not\in \Lambda_{[T]}$, then there would be at least one component index $n$ such that $\mu^{(n)}_{[T]} > \mu_{[T]}'^{(n)}$ (${\boldsymbol \mu}'_{[T]} \in \overline{\mathcal{M}}_{[T]}$).
Thus, $\forall {\boldsymbol \mu}'_{[T]} \in \overline{\mathcal{M}}_{[T]}$,
\begin{align}
\underset{n\in\mathcal{N}} {\min} \left\{\frac{\mu'^{(n)}_{[T]}}{\mu_{[T]}^{(n)}}\right\} < 1,
\end{align}
which implies $\delta_T({\boldsymbol \mu}_{[T]}) < 1$, and this contradicts $\delta_T({\boldsymbol \mu}_{[T]}) \geq 1$.
Therefore, ${\boldsymbol \mu}_{[T]} \in \Lambda_{[T]}$.
\end{IEEEproof}

%
%

Another important property of rate margin is that it quantifies the extent of which an achievable rate-tuple can be linearly scaled-up while remaining achievable.

\begin{proposition}\label{prop:Achievable Rate's Maximum Magnification for Remaining Achievable}
$\forall {\boldsymbol \mu}_{[T]} \in \Lambda_{[T]}$, the rate margin gives the maximum scalar $\rho$ such that $\rho{\boldsymbol \mu}_{[T]}$ is still an achievable rate-tuple, i.e.,
\begin{align}\label{eqn:Achievable Rate's Maximum Magnification for Remaining Achievable}
\underset {\rho{\boldsymbol \mu}_{[T]} \in \Lambda_{[T]}} {\max} \rho = \delta_T({\boldsymbol \mu}_{[T]}) \geq 1.
\end{align}
\end{proposition}

\begin{IEEEproof}
For convenience, we label $\rho^* = \underset {\rho{\boldsymbol \mu}_{[T]} \in \Lambda_{[T]}} {\max} \rho$ for the left-hand-side of~\eqref{eqn:Achievable Rate's Maximum Magnification for Remaining Achievable}.
Let ${\boldsymbol \mu}'_{[T]} = \rho^*{\boldsymbol \mu}_{[T]}$, and we have ${\boldsymbol \mu}'_{[T]} \in \Lambda_{[T]}$.
In addition, by setting ${\boldsymbol \mu}''_{[T]} = \delta_T({\boldsymbol \mu}_{[T]}){\boldsymbol \mu}_{[T]}$, the proof starts as follows.

i).~$\delta_T({\boldsymbol \mu}_{[T]}) \geq \rho^*$: assume $\delta_T({\boldsymbol \mu}_{[T]}) < \rho^*$, then $\delta_T({\boldsymbol \mu}''_{[T]}) < \delta_T({\boldsymbol \mu}'_{[T]}) \leq 1$.
It implied $\delta_T({\boldsymbol \mu}''_{[T]}) < 1$, which contradicts to the following derivation
\begin{multline}\label{eqn:Proof Equation 1 for Achievable Rate's Maximum Magnification for Remaining Achievable}
\delta_T({\boldsymbol \mu}''_{[T]}) = \underset{{\boldsymbol \mu}'''_{[T]} \in \overline{\mathcal{M}}_{[T]}~~}{\max} \underset{n\in\mathcal{N}} {\min} \left\{\frac{\mu_{[T]}'''^{(n)}}{\mu''^{(n)}_{[T]}}\right\} \\= \frac {1} {\delta_T({\boldsymbol \mu}_{[T]})} \underset{{\boldsymbol \mu}'''_{[T]} \in \overline{\mathcal{M}}_{[T]}~~}{\max} \underset{n\in\mathcal{N}} {\min} \left\{\frac{\mu_{[T]}'''^{(n)}}{\mu''^{(n)}_{[T]}}\right\} = 1.
\end{multline}
Therefore, $\delta_T({\boldsymbol \mu}_{[T]}) \geq \rho^*$ holds.

ii).~$\delta_T({\boldsymbol \mu}_{[T]}) \leq \rho^*$: assume $\delta_T({\boldsymbol \mu}_{[T]}) > \rho^*$, and the proof is similar to that in i) (equation~\eqref{eqn:Proof Equation 1 for Achievable Rate's Maximum Magnification for Remaining Achievable} still holds).

To sum up, $\delta_T({\boldsymbol \mu}_{[T]}) = \rho^*$.
According to \propref{prop:Completeness of Rate Margin}, $\delta_T({\boldsymbol \mu}_{[T]}) \geq 1$ and~\eqref{eqn:Achievable Rate's Maximum Magnification for Remaining Achievable} is satisfied.
\end{IEEEproof}

With \propref{prop:Achievable Rate's Maximum Magnification for Remaining Achievable}, we can also derive an important property that the rate margin quantifies the extent of which an unachievable rate-tuple should be linearly scaled down to become achievable.

\begin{proposition}\label{prop:Unachievable Rate's Minimum Minification for Becoming Achievable}
$\forall {\boldsymbol \mu}_{[T]} \not\in \Lambda_{[T]}$, the rate margin gives the minimum scalar $r$ such that ${\boldsymbol \mu}_{[T]}/r$ becomes an achievable rate-tuple, i.e.,
\begin{align}\label{eqn:Unachievable Rate's Minimum Minification for Becoming Achievable}
\underset {{\boldsymbol \mu}_{[T]}/r \in \Lambda_{[T]}} {\min} r = \frac {1} {\delta_T({\boldsymbol \mu}_{[T]})} > 1.
\end{align}
\end{proposition}

\begin{IEEEproof}
Since the proof is similar to that in~\propref{prop:Achievable Rate's Maximum Magnification for Remaining Achievable}, we omit it here.
\end{IEEEproof}

Last but not least, the rate margin is an indictor for those rate-tuples on the weak Pareto frontier, which is also very useful for results to be derived later.

\begin{corollary}\label{cor:Boundary Rate Completeness of Rate Margin}
${\boldsymbol \mu}_{[T]} \in \mathcal{M}_{[T]}$ if and only if $\delta_T({\boldsymbol \mu}_{[T]}) = 1$.
\end{corollary}

\begin{IEEEproof}
Based on \propref{prop:Achievable Rate's Maximum Magnification for Remaining Achievable} and \propref{prop:Unachievable Rate's Minimum Minification for Becoming Achievable}, the proof is straightforward.
\end{IEEEproof}

\begin{figure}[h]
\centering
\includegraphics [width=0.6\columnwidth]{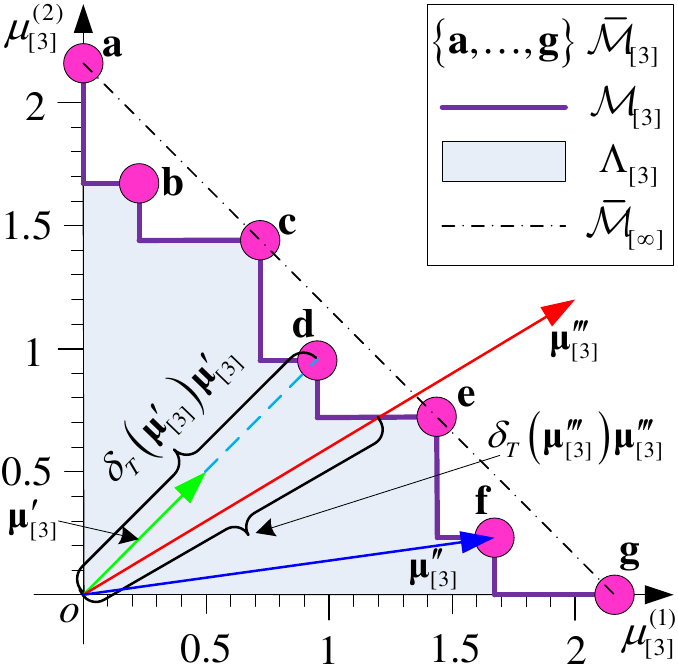}
\caption{Illustration of rate margin.
The network parameters are the same as those in \figref{fig:Example for Throughput Region 3}, but the chosen rate-tuples are different:
${\boldsymbol \mu}'_{[3]} = [0.5, 0.5]^{\mathrm{T}}$, ${\boldsymbol \mu}''_{[3]} = [1.6729, 0.2316]^{\mathrm{T}}$, and ${\boldsymbol \mu}'''_{[3]} = [2, 1.2]^{\mathrm{T}}$.
${\boldsymbol \mu}'_{[3]}$ is in $\Lambda_{[3]}\setminus\mathcal{M}_{[3]}$ (i.e., the interior of $\Lambda_{[3]}$), and ${\boldsymbol \mu}''_{[3]}$ is in $\mathcal{M}_{[3]}$, but ${\boldsymbol \mu}'''_{[3]}$ is not in $\Lambda_{[3]}$.
The rate margins are $\delta_3({\boldsymbol \mu}'_{[3]}) = 1.9046$, $\delta_3({\boldsymbol \mu}''_{[3]}) = 1$, and $\delta_3({\boldsymbol \mu}'''_{[3]}) = 0.6006$.
}
\label{fig:Pictorial Illustration on Rate Margin}
\end{figure}

\begin{remark}[Properties of Rate Margin]\label{rek:Summary of Properties of Rate Margin}
We use a numerical example to summarize the properties of rate margin. Specifically, we test three rate-tuples, i.e., ${\boldsymbol \mu}'_{[3]}$, ${\boldsymbol \mu}''_{[3]}$ and ${\boldsymbol \mu}'''_{[3]}$, whose values are shown in the caption of \figref{fig:Pictorial Illustration on Rate Margin}.
Using \propref{prop:Completeness of Rate Margin}, we have $\delta_3({\boldsymbol \mu}'_{[3]}) \geq 1$, $\delta_3({\boldsymbol \mu}''_{[3]}) \geq 1$, and $\delta_3({\boldsymbol \mu}'''_{[3]}) < 1$, hence we know that ${\boldsymbol \mu}'_{[3]} \in \Lambda_{[3]}$, ${\boldsymbol \mu}''_{[3]} \in \Lambda_{[3]}$, and ${\boldsymbol \mu}'''_{[3]} \not\in \Lambda_{[3]}$.
These conclusions are correct as shown in \figref{fig:Pictorial Illustration on Rate Margin}.
\propref{prop:Achievable Rate's Maximum Magnification for Remaining Achievable} implies that ${\boldsymbol \mu}'_{[3]}$ can be linearly scaled up by at most $\delta_3({\boldsymbol \mu}'_{[3]}) = 1.9046$ and remains achievable.
Similarly, \propref{prop:Unachievable Rate's Minimum Minification for Becoming Achievable} tells that ${\boldsymbol \mu}'''_{[3]}$ should be linearly scaled down by at least $\delta_3({\boldsymbol \mu}'''_{[3]}) = 0.6006$ and becomes achievable.
By \corref{cor:Boundary Rate Completeness of Rate Margin}, $\delta_3({\boldsymbol \mu}''_{[3]}) = 1$ (hence ${\boldsymbol \mu}''_{[3]} \in \mathcal{M}_{[3]}$), and there is no room for ${\boldsymbol \mu}''_{[3]}$ to be linearly scaled up and remains achievable.
\end{remark}

Now, we give the definition of rate-achieving policy.

\begin{definition}[Rate-Achieving Policy]\label{def:Rate-Achieving Policy}
For a given transmit-power-tuple set $\mathcal{S}$ and $T$ time slots, $\forall {\boldsymbol \mu}_{[T]} \in \Lambda_{[T]}$, the rate-achieving policy for ${\boldsymbol \mu}_{[T]}$ is a sequence of rate-power pairs
\begin{align}\label{eqn:Rate-Achieving Policy}
\mathcal{P}_{T} = \left({\boldsymbol \mu}_t, \mathbf{s}_t\right)_{t=1}^{T},\quad\mathbf{s}_t \in \mathcal{S},
\end{align}
with the maximum rate constraints
\begin{align}\label{eqn:Capacity Constraints}
{\boldsymbol \mu}_t \preceq {\boldsymbol \mu}_{\max}(\mathbf{s}_t, L, \epsilon),
\end{align}
such that ${\boldsymbol \mu}_{[T]}$ can be achieved, i.e.,
\begin{align}\label{eqn:Rate Achievement Equation}
{\boldsymbol \mu}_{[T]} = \frac{1} {T} \sum_{t=1}^{T} {\boldsymbol \mu}_t.
\end{align}
\end{definition}

Equation~\eqref{eqn:Capacity Constraints} means that: for each transmitter-receiver pair, the transmission rate should not exceed the corresponding maximum rate.

\begin{remark}
\defref{def:Rate-Achieving Policy} tells that if ${\boldsymbol \mu}_{[T]} \in \Lambda_{[T]}$ (or equivalently $\delta_T({\boldsymbol \mu}_{[T]}) \geq 1$, see \propref{prop:Completeness of Rate Margin}), then there always exist some policies $\mathcal{P}_T$ described by~\eqref{eqn:Rate-Achieving Policy},~\eqref{eqn:Capacity Constraints}, and~\eqref{eqn:Rate Achievement Equation} to achieve ${\boldsymbol \mu}_{[T]}$.
To be more specific, a power sequence $(\mathbf{s}_t)_{t=1}^{T}$ provides the maximum rates to support the rate-tuple ${\boldsymbol \mu}_{[T]}$ (see~\eqref{eqn:Capacity Constraints}), which is eventually achieved by rate-tuple sequence $({\boldsymbol \mu}_t)_{t=1}^{T}$ (see~\eqref{eqn:Rate Achievement Equation}).
The existence of $\mathcal{P}_T$ is exactly guaranteed by~\eqref{eqn:Finite-Horizon Throughput Region} in \defref{def:Finite-Slots Throughput Region} and~\eqref{eqn:One-Slot Throughput Region} that: $\forall {\boldsymbol \mu}_t \in \Lambda_{[1]}$, there at least exists one ${\boldsymbol \mu}_{\max}(\mathbf{s}_t, L, \epsilon)$ such that ${\boldsymbol \mu}_t \in {\boldsymbol \mu}_{\max}(\mathbf{s}_t, L, \epsilon)$.
Nevertheless, \defref{def:Rate-Achieving Policy} just only states the existence of $\mathcal{P}_T$ for an achievable rate-tuple, but how to find an efficient algorithm to find $\mathcal{P}_T$ is yet to be investigated.
\end{remark}

After defining the rate margin and rate-achieving policy, we now focus on two key problems
\begin{itemize}
\item How to efficiently compute the rate margin $\delta_T({\boldsymbol \mu}_{[T]})$.
\item If $\delta_T({\boldsymbol \mu}_{[T]}) \geq 1$, how to design a rate-achieving policy with high computational efficiency.
\end{itemize}


\section{Derivation of Rate Margin and Rate-Achieving Policy}\label{sec:Derivation of Rate Margin and Rate-Achieving Policy}

Although deriving the rate margin and rate-achieving policy are two different problems, we carefully formulate them into the following joint problem (see \probref{prob:Rate-Margin-Deriving Problem}) from the viewpoint of data transmission.
Note that achieving a given average rate-tuple ${\boldsymbol \mu}_{[T]}$ over $T$ time slots is the same as transmitting $T{\boldsymbol \mu}_{[T]} L$ amount of data from data queues within $T$ time slots (recall that $L$ is the blocklength).

\begin{problem}\label{prob:Rate-Margin-Deriving Problem}
For a given transmit-power-tuple set $\mathcal{S}$ and $T$ time slots, we consider the following problem:
\begin{align}\label{eqn:Rate-Margin-Deriving Problem}
\begin{array}{c l}
\underset {\left(\mathbf{s}_t\right)_{t=1}^{p},\,\mathbf{s}_t\in\mathcal{S}} {\mathrm{minimize}} & p + \underset {n\in\mathcal{N}} {\max} \left[\displaystyle\frac {Q_0^{(n)}} {\sum_{t=1}^{p} \mu_{\max}^{(n)}(\gamma_n(\mathbf{s}_t), L, \epsilon)L}\right]\\
\mathrm{subject~to}~&\mathbf{Q}_{t} = \left(\mathbf{Q}_{t-1} - {\boldsymbol \mu}_{\max}(\mathbf{s}_t, L, \epsilon) L\right)^+,\\
&\mathbf{Q}_{0} = T {\boldsymbol \mu}_{[T]} L,\\
&\mathbf{Q}_{p-1} \neq \mathbf{0},\\
&\mathbf{Q}_{p} = \mathbf{0},
\end{array}
\end{align}
where $t\in\{1,\dotsc,p\}$.
The data queues are $\mathbf{Q}_t = [Q_t^{(1)},\dotsc,Q_t^{(N)}]^{\mathrm{T}}$, each $Q_t^{(n)} \in \overline{\mathbb{R}}_+$ is the length of the data queue for transmitter $n$ after $\mathbf{s}_t$ is applied in time slot $t$ ($t \in \{1,\dotsc,p\}$).
In addition, $\mathbf{Q}_0 = T {\boldsymbol \mu}_{[T]} L$ is the initial lengths of queues before applying $\mathbf{s}_1$, which means the given rate-tuple ${\boldsymbol \mu}_{[T]} = [\mu_{[T]}^{(1)},\dotsc,\mu_{[T]}^{(N)}]^{\mathrm{T}}$ at which transmission must take place in order to send a total of $T {\boldsymbol \mu}_{[T]} L$ amount of data, and $L$ is the blocklength.
Note that $p$ denotes the number of time slots for transmission (since $\mathbf{Q}_{p-1} \neq \mathbf{0}$ and $\mathbf{Q}_{p} = \mathbf{0}$) and is a variable dependent on $\left(\mathbf{s}_t\right)_{t=1}^{p}$.
The optimal solution (not unique for $T > 1$) is denoted as $(\mathbf{s}_t^*)_{t=1}^{p^*}$.
The corresponding data-queues sequence under optimal solution is $(\mathbf{Q}_t^*)_{t=1}^{p^*}$.
The optimal objective is
\begin{align}\label{eqn:Optimal Objective}
p^* + \underset {n\in\mathcal{N}} {\max} \left[\frac {Q_0^{(n)}} {\sum_{t=1}^{p^*} \mu_{\max}^{(n)}(\gamma_n(\mathbf{s}_t^*), L, \epsilon)L}\right].
\end{align}
\end{problem}

The following two lemmas explain the meaning of the first and second items in~\eqref{eqn:Optimal Objective}, respectively.

\begin{lemma}\label{lem:Minimum Transmission Time Slots}
$p^*$ in~\eqref{eqn:Optimal Objective} is the minimum number of time slots that clears the data queues $\mathbf{Q}_0$.
\end{lemma}

\begin{IEEEproof}
By~\eqref{eqn:Rate-Margin-Deriving Problem}, $p$ is the number of time slots that clears $Q_0^{(n)}$, which implies the following
\begin{align}\label{eqn:The Last Term of Objective is always not greater than 1}
\underset {n\in\mathcal{N}} {\max} \left[\frac {Q_0^{(n)}} {\sum_{t=1}^{p} \mu_{\max}^{(n)}(\gamma_n(\mathbf{s}_t), L, \epsilon)L}\right] \leq 1.
\end{align}
Then $p^*$ in~\eqref{eqn:Optimal Objective} is the optimal number of time slots to clear $Q_0^{(n)}$.
This is because we can never find a $p' < p^*$ such that
\begin{multline}
p' + \underset {n\in\mathcal{N}} {\max} \left[\frac {Q_0^{(n)}} {\sum_{t=1}^{p'} \mu_{\max}^{(n)}(\gamma_n(\mathbf{s}_t'^*), L, \epsilon)L}\right]
\\\geq p^* + \underset {n\in\mathcal{N}} {\max} \left[\frac {Q_0^{(n)}} {\sum_{t=1}^{p^*} \mu_{\max}^{(n)}(\gamma_n(\mathbf{s}_t^*), L, \epsilon)L}\right]
\end{multline}
holds with~\eqref{eqn:The Last Term of Objective is always not greater than 1} (since $p'$ and $p^*$ are integers).
\end{IEEEproof}

The value of $p^*$ tells the rate-achievability:
If $p^* \leq T$, then it is possible to transmit $T{\boldsymbol \mu}_{[T]}L$ amount of data within $T$ slots.
In other words, the rate-tuple ${\boldsymbol \mu}_{[T]}$ is achievable within $T$ slots.
If $p^* > T$, then it implies the rate-tuple ${\boldsymbol \mu}_{[T]}$ is unachievable within $T$ slots.

\begin{lemma}\label{lem:Deriving Rate Margin when p^* = T}
The rate margin for ${\boldsymbol \mu}_{[p^*]}$ in $\Lambda_{[p^*]}$ is the reciprocal of the second item in~\eqref{eqn:Optimal Objective}, i.e.,
\begin{align}\label{eqn:Deriving Rate Margin when p^* = T}
\delta_{p^*}({\boldsymbol \mu}_{[p^*]}) = \frac {1} {\underset {n\in\mathcal{N}} {\max} \left[\frac {Q_0^{(n)}} {\sum_{t=1}^{p^*} \mu_{\max}^{(n)}(\gamma_n(\mathbf{s}_t^*), L, \epsilon)L}\right]}.
\end{align}
\end{lemma}

\begin{IEEEproof}
With the definition of the rate margin (see \defref{def:Rate Margin}) and replacing $T$ with $p^*$, we can derive
\begin{equation}\label{eqn:Reciprocal for Rate Margin 1}
\begin{split}
\frac {1} {\delta_{p^*}({\boldsymbol \mu}_{[p^*]})} &\stackrel{(a)}{=} \underset{{\boldsymbol \mu}'_{[p^*]} \in \overline{\mathcal{M}}_{[p^*]}~~}{\min} \underset{n\in\mathcal{N}} {\max} \left\{\frac{\mu^{(n)}_{[p^*]}}{\mu'^{(n)}_{[p^*]}}\right\}\\
&= \underset{{\boldsymbol \mu}'_{[p^*]} \in \overline{\mathcal{M}}_{[p^*]}~~}{\min} \underset{n\in\mathcal{N}} {\max} \left\{\frac{p^*\mu^{(n)}_{[p^*]}L}{p^*\mu'^{(n)}_{[p^*]}L}\right\},
\end{split}
\end{equation}
where $(a)$ follows from that the $\max$ ($\min$) of ${\mu'^{(n)}_{[p^*]}} / {\mu^{(n)}_{[p^*]}}$ is the $\min$ ($\max$) of ${\mu^{(n)}_{[p^*]}}/{\mu'^{(n)}_{[p^*]}}$.
Since $p^*\mu^{(n)}_{[p^*]}L = \mathbf{Q}_0^{(n)}$, there exists $(\mathbf{s}_t)_{t=1}^{p^*}$ such that
\begin{align}
p^*\mu'^{(n)}_{[p^*]} L = \sum_{t=1}^{p^*} \mu'^{(n)}_{t} L \stackrel {(b)} {=} \sum_{t=1}^{p^*} \mu_{\max}^{(n)}(\gamma_n(\mathbf{s}_t), L, \epsilon) L,
\end{align}
where $(b)$ holds for ${\boldsymbol \mu}'_{[p^*]} \in \overline{\mathcal{M}}_{[p^*]}$.
Thus, we rewrite~\eqref{eqn:Reciprocal for Rate Margin 1} as
\begin{align}\label{eqn:Reciprocal for Rate Margin 2}
\begin{split}
\!\!\frac {1} {\delta_{p^*}({\boldsymbol \mu}_{[p^*]})} &= \underset{\left(\mathbf{s}_t\right)_{t=1}^{p^*}~}{\min} \underset{n\in\mathcal{N}} {\max} \left[\frac{Q_0^{(n)}}{\sum_{t=1}^{p^*} \mu_{\max}^{(n)}(\gamma_n(\mathbf{s}_t), L, \epsilon)L}\right] \\&\stackrel {(c)}{=} \max_{n\in\mathcal{N}} \left[\frac{Q_0^{(n)}}{\sum_{t=1}^{p^*} \mu_{\max}^{(n)}(\gamma_n(\mathbf{s}_t^*), L, \epsilon)L}\right],
\end{split}
\end{align}
where $(c)$ follows from \lemref{lem:Minimum Transmission Time Slots} and the objective in~\eqref{eqn:Rate-Margin-Deriving Problem}.
%
Therefore,~\eqref{eqn:Deriving Rate Margin when p^* = T} holds.
\end{IEEEproof}

Note that the reciprocal of the second item in~\eqref{eqn:Optimal Objective} is the rate margin for $p^*$ time slots.
However, we want to derive the rate margin for $T$ time slots ($p^*$ is not necessarily equal to $T$).
In the rest of this section, we discuss how to derive the rate margin for any given $T$ time slots (see \secref{sec:Deriving Rate Margin}) based on the optimal objective of \probref{prob:Rate-Margin-Deriving Problem}.
Furthermore, by the optimal solution of \probref{prob:Rate-Margin-Deriving Problem} the rate-achieving policy is derived (see \secref{sec:Deriving Rate-Achieving Policy}).
Finally, an efficient solution method of \probref{prob:Rate-Margin-Deriving Problem} is given in \secref{sec:Solution for Rate-Margin-Deriving Problem}.

\subsection{Deriving Rate Margin}\label{sec:Deriving Rate Margin}

This subsection proposes a method to derive the rate margin by iteratively solving \probref{prob:Rate-Margin-Deriving Problem}.

Firstly, if we find $p^* = T$ after solving \probref{prob:Rate-Margin-Deriving Problem}, then the rate margin can be derived directly from \lemref{lem:Deriving Rate Margin when p^* = T}, since $\delta_{T}({\boldsymbol \mu}_{[T]}) = \delta_{p^*}({\boldsymbol \mu}_{[p^*]})$ in this case.

If $p^* \neq T$, we can use an iteration strategy to derive rate margin $\delta_{T}({\boldsymbol \mu}_{[T]})$.
To distinguish $p^*$ (and $\mathbf{Q}_0$) in different iterations, we label $p^*_k$ (and $\mathbf{Q}_{0,k}$) as the $p^*$ (and $\mathbf{Q}_0$) for the $k$\textsuperscript{th} iteration.
Now, the main idea of our iteration strategy is given as follows:
based on the information from \lemref{lem:Deriving Rate Margin when p^* = T}, we linearly scale the initial condition $\mathbf{\mathbf{Q}}_{0,k}$ in \probref{prob:Rate-Margin-Deriving Problem} for each iteration $k \in \{1,\dotsc,K\}$, until $p^*_K = T$, in which case, $\delta_{T}({\boldsymbol \mu}_{[T]}) = \delta_{p^*_K}({\boldsymbol \mu}_{[p^*_K]})$.
As such, the rate margin can be finally determined recursively in a finite number of steps whenever $K$ is finite.
The iteration strategy is given in \algref{alg:Deriving Rate Margin}, and proved in \thmref{thm:Calculation of Rate Margin}.

\begin{algorithm}
\begin{footnotesize}
\caption{Deriving Rate Margin}\label{alg:Deriving Rate Margin}
\begin{algorithmic}[1]
\REQUIRE
    $T$: the number of time slots; $N$: the number of transmitter-receiver pairs; ${\boldsymbol \mu}_{[T]}$: the given average rate-tuple;
    $\mathcal{S}$: the transmit-power-tuple set; $L$: the blocklength; $\epsilon$: the error probability.
\ENSURE
    $\delta_T({\boldsymbol \mu}_{[T]})$: the rate margin.
\STATE  Initialization: $k = 1$; $\mathbf{Q}_{0, k} = T {\boldsymbol \mu}_{[T]} L$; $p_k^* = 0$; $\mathrm{flag} = 0$; $\varepsilon = 10^{-7}$ \COMMENT{\textbf{comments:} $\varepsilon$ is the precision for rate-tuple calculation}.
\WHILE  {$p_k^* \neq T$}
    \STATE  Solve \probref{prob:Rate-Margin-Deriving Problem} with $\mathbf{Q}_{0} = \mathbf{Q}_{0,k}$, and derive $\delta_{p^*_{k}}({\boldsymbol \mu}_{[p^*_{k}],k})$ by \lemref{lem:Deriving Rate Margin when p^* = T};\label{line:Solve Problem 1}
    \IF {$p_k^* < T$}
        \STATE  $\mathbf{Q}_{0,k+1} = \mathbf{Q}_{0,k} \delta_{p^*_{k}}({\boldsymbol \mu}_{[p^*_{k}],k}) \lfloor {T} / {p^*_{k}}\rfloor + R_k \rho {\boldsymbol \mu}_{[T]}$, where $R_k := T~\mathrm{mod}~p^*_k$, and $\rho = \delta_1({\boldsymbol \mu}_{[T]})$; $\mathrm{flag} = 1$;\label{line:a}
        \IF {$\lfloor {T}/{p^*_{k}}\rfloor == 1$ and $\rho == 0$}
            \STATE  $\mathbf{Q}_{0,k+1} = \mathbf{Q}_{0,k} + \varepsilon T$;\label{line:middle a}
        \ENDIF\label{line:end a}
    \ELSIF  {$p_k^* > T$ and $\mathrm{flag} == -1$}
        \STATE  $\mathbf{Q}_{0,k+1} = 0$; $p_k^* = T$ \COMMENT{\textbf{comments:} condition for ending the loop};\label{line:b}
    \ELSIF  {$p_k^* > T$ and $\mathrm{flag} == 0$}
        \STATE  $\mathbf{Q}_{0,k+1} = T \max\{\rho, \varepsilon\}{\boldsymbol \mu}_{[T]}$, where $\rho = \delta_1({\boldsymbol \mu}_{[T]})$; $\mathrm{flag} = -1$;\label{line:c}
    \ELSIF  {$p_k^* > T$ and $\mathrm{flag} == 1$}
        \STATE  $\mathbf{Q}_{0,k+1} = \mathbf{Q}_{0,k} - \varepsilon T$; $p_k^* = T$;\label{line:d}
    \ELSE
        \STATE  $\mathbf{Q}_{0,k+1} = \mathbf{Q}_{0,k} \delta_{p^*_{k}}({\boldsymbol \mu}_{[p^*_{k}],k})$; \COMMENT{\textbf{comments:} $p_k^* = T$} \label{line:e}
    \ENDIF
    \STATE  $K = k$; $k = k + 1$; \COMMENT{\textbf{comments:} $K$ is the total iteration number.}
\ENDWHILE
\RETURN $\delta_T({\boldsymbol \mu}_{[T]}) = Q_{0,K+1}^{(1)}/Q_{0,1}^{(1)}$.\label{line:Deriving Rate Margin}
\end{algorithmic}
\end{footnotesize}
\end{algorithm}

\begin{theorem}[Calculation of Rate Margin]\label{thm:Calculation of Rate Margin}
For ${\boldsymbol \mu}_{[T]} \succ 0$,\footnote{\algref{alg:Deriving Rate Margin} is valid for rate-tuple ${\boldsymbol \mu}_{[T]}$ whose components are all greater than $0$. This is without loss of generality because the zero components stand for zero transmissions, and hence we can remove these inactive transmitter-receiver pairs from the network model.} the rate margin can be obtained by \algref{alg:Deriving Rate Margin} involving a finite number of $K$ iterations, upper bounded by
\begin{align}\label{eqn:Upper Bound on K}
K \leq
\begin{cases}
T - p_1^* + 1 & p_1^* < T,\\
1 & p_1^* = T,\\
T - p_2^* + 2 & p_1^* > T.
\end{cases}
\end{align}
\end{theorem}

\begin{IEEEproof}
See \apxref{apx:Proof of Theorem 1}.
\end{IEEEproof}

\begin{remark}
The inequality in~\eqref{eqn:Upper Bound on K} gives an upper bound on the number of iterations.
For example, considering the case $p_1^* < T$, we have $K \leq T - p_1^* + 1$, which means that the rate margin requires solving \probref{prob:Rate-Margin-Deriving Problem} at most $T - p_1^* + 1$ times.
\end{remark}

Before closing this subsection, we give a useful corollary.
The proof can be easily obtained by \propref{prop:Completeness of Rate Margin} and the proof of \thmref{thm:Calculation of Rate Margin}.

\begin{corollary}\label{cor:Three Equivalent Statements}
The following three statements are equivalent: i)~${\boldsymbol \mu}_{[T]} \in \Lambda_{[T]}$; ii)~$\delta_T({\boldsymbol \mu}_{[T]}) \geq 1$; iii)~$p^*_1 \leq T$.
\end{corollary}

\subsection{Deriving Rate-Achieving Policy}\label{sec:Deriving Rate-Achieving Policy}

In this subsection, we derive a rate-achieving policy for any given achievable rate-tuple.
It should be noted that our method is complete, i.e., for any given achievable rate-tuple, the corresponding rate-achieving policy can be obtained.

We present a rate-achieving policy for all rate-tuples in the $T$-slot throughput region as follows.

\begin{theorem}[Rate-Achieving Policy for All Achievable Rates]\label{thm:Rate-Achieving Policy for All Achievable Rates}
Given a transmit-power-tuple set $\mathcal{S}$ and a finite horizon of $T$ time slots, then:
\begin{enumerate}[i)]
\item   If ${\boldsymbol \mu}_{[T]} \in \Lambda_{[T]}$, then $p^* \leq T$, and the rate-achieving policy is $\mathcal{P}_T = ({\boldsymbol \mu}_t, \mathbf{s}_t)_{t=1}^T$ with
\begin{align}\label{eqn:Rate-Achieving Policy for All Achievable Rates}
\left({\boldsymbol \mu}_t, \mathbf{s}_t\right) =
\begin{cases}
\left(\frac {\mathbf{Q}_{t-1}^* - \mathbf{Q}_{t}^*} {L}, \mathbf{s}_t^*\right)& 1 \leq t \leq p^*,\\
\left(\mathbf{0}, \mathbf{0}\right)& p^* < t \leq T,
\end{cases}
\end{align}
where $(\mathbf{s}_t^*)_{t=1}^{p^*}$, is an optimal solution to \probref{prob:Rate-Margin-Deriving Problem} and $\mathbf{Q}_t^*$ is the corresponding data queue vector in time slot $t$ when applying the optimal solution.
\item   If ${\boldsymbol \mu}_{[T]} \not\in \Lambda_{[T]}$, then solving \probref{prob:Rate-Margin-Deriving Problem} gives $p^* > T$.
\end{enumerate}
\end{theorem}

\begin{IEEEproof}
See \apxref{apx:Proof of Theorem 2}.
\end{IEEEproof}

\subsection{Solution for \probref{prob:Rate-Margin-Deriving Problem}}\label{sec:Solution for Rate-Margin-Deriving Problem}

In \secref{sec:Deriving Rate Margin} and \secref{sec:Deriving Rate-Achieving Policy}, all main results are based on the solution of \probref{prob:Rate-Margin-Deriving Problem}.
Therefore, designing an efficient algorithm to solve this problem can directly improve the efficiency of deriving rate margin and rate-achieving policy.
In this subsection, we discuss how to efficiently solve \probref{prob:Rate-Margin-Deriving Problem}.

To solve~\eqref{eqn:Rate-Margin-Deriving Problem} in \probref{prob:Rate-Margin-Deriving Problem}, intuitively, we could use dynamic programming to search from $\mathbf{Q}_p = \mathbf{0}$ to $\mathbf{Q}_0 = T{\boldsymbol \mu}_{[T]} L$ (backwards) or employ other uninformed search strategies~\cite{RussellS2009BOOK}.
However, in such searching methods, the complexity is $O(|\mathcal{S}|^{p^*})$, where $\mathcal{S}$ is the transmit-tuple set, and $p^*$ is the minimum transmission time (see \lemref{lem:Minimum Transmission Time Slots}) which can be larger than $T$.
%

For example, if we start the search from $\mathbf{Q}_0 = T {\boldsymbol \mu}_{[T]} L$, for the first step, we will calculate all possible
\begin{align}\label{eqn:Example for The Leaf Nodes in Depth 1}
\mathbf{Q}_1 = \left(\mathbf{Q}_0 - {\boldsymbol \mu}_{\max}(\mathbf{s}_t, L, \epsilon) L \right)^+,
\end{align}
for all $\mathbf{s}_1 \in \mathcal{S}$.
Thus, the number of leaf nodes is $|\mathcal{S}|$ for the depth $t = 1$.
Similarly, for every $\mathbf{Q}_1$ in~\eqref{eqn:Example for The Leaf Nodes in Depth 1}, we have $|\mathcal{S}|$ possible $\mathbf{Q}_2$, and thus the leaf nodes for $t = 2$ is $|\mathcal{S}|^2$.
As such, the number of leaf nodes for depth $t = p^*$ (since the optimal transmission time is $p^*$, see \lemref{lem:Minimum Transmission Time Slots}, and we need compare all the objective functions in this depth) is $|\mathcal{S}|^{p^*}$.
Thus, the complexity of such searching methods is $O(|\mathcal{S}|^{p^*})$.

In this subsection, we use the following three steps to significantly improve the computational efficiency in solving \probref{prob:Rate-Margin-Deriving Problem}.
The resulting complexity is $O(B^{\min\{p^*, T\}})$, where $B$ (the effective branching factor\footnote{It is a very popular metric for characterizing the efficiency of a searching method, see Section 3.6.1 in~\cite{RussellS2009BOOK}.}) is a much smaller number compared to $|\mathcal{S}|$, and $p^*$ is reduced to $\min\{p^*, T\}$ for the case $p^* > T$.

For the convenience of applying our search algorithm, we modify the objective in~\eqref{eqn:Rate-Margin-Deriving Problem} as
\begin{align}\label{eqn:Modified Objective in Rate-Margin-Derving Problem}
p - 1 + \underset {n\in\mathcal{N}} {\max} \left[\frac {Q_0^{(n)}} {\sum_{t=1}^{p} \mu_{\max}^{(n)}(\gamma_n(\mathbf{s}_t), L, \epsilon)L}\right],
\end{align}
by adding $-1$ to the original objective.
It is readily to see that this modification does not affect the optimal solution (i.e., the original and modified objectives have the same optimal solution).

\textit{Step 1}: Firstly, we reduce the branching factor from $|\mathcal{S}|$ to $|\overline{\mathcal{M}}_{[1]}|$, which is given in \propref{prop:Optimization for the Branching Factor}.
%
%
%
%
%
%
%

\begin{proposition}\label{prop:Optimization for the Branching Factor}
There exists a sequence $(\mathbf{s}_t)_{t=1}^{p^*}$, where ${\boldsymbol \mu}_{\max}(\mathbf{s}_t, L, \epsilon) \in \overline{\mathcal{M}}_{[1]},\,t\in\{1,\dotsc,p^*\}$, such that $(\mathbf{s}_t)_{t=1}^{p^*}$ itself is an optimal solution of \probref{prob:Rate-Margin-Deriving Problem}.
\end{proposition}

\begin{IEEEproof}
Let $(\mathbf{s}^*_t)_{t=1}^{p^*}$ be any optimal solution of \probref{prob:Rate-Margin-Deriving Problem}, we have $\mathbf{Q}_{p^*} = \mathbf{0}$, which implies
\begin{align}\label{eqn:Data Cleared Equation}
T {\boldsymbol \mu}_{[T]} L \preceq \sum_{t=1}^{p^*} {\boldsymbol \mu}_{\max}(\mathbf{s}_t, L, \epsilon) L.
\end{align}
Let $(\mathbf{s}_t)_{t=1}^{p^*}$ be the sequence that ${\boldsymbol \mu}_{\max}(\mathbf{s}_t, L, \epsilon) \in \overline{\mathcal{M}}_{[1]},\,t\in\{1,\dotsc,p^*\}$, and ${\boldsymbol \mu}_{\max}(\mathbf{s}_t^*, L, \epsilon) \preceq {\boldsymbol \mu}_{\max}(\mathbf{s}_t, L, \epsilon)$.
Thus,~\eqref{eqn:Data Cleared Equation} can be rewritten as
\begin{align}
T {\boldsymbol \mu}_{[T]} L \preceq \sum_{t=1}^{p^*} {\boldsymbol \mu}_{\max}(\mathbf{s}_t^*, L, \epsilon) L \preceq \sum_{t=1}^{p^*} {\boldsymbol \mu}_{\max}(\mathbf{s}_t, L, \epsilon) L,
\end{align}
which implies $\mathbf{Q}_{p^*} = \mathbf{0}$ when applying $(\mathbf{s}_t)_{t=1}^{p^*}$.
Therefore, $(\mathbf{s}_t)_{t=1}^{p^*}$ is an optimal solution of \probref{prob:Rate-Margin-Deriving Problem}.
\end{IEEEproof}

\begin{remark}\label{rek:Reduction in the Branching Factor}
\propref{prop:Optimization for the Branching Factor} tells that we only need to consider the transmit powers corresponding to the rate-tuple on the Pareto frontier of the $1$-slot throughput region, instead of all possible transmit powers. Hence, the transmit-power-tuple set $\mathcal{S}$ in \probref{prob:Rate-Margin-Deriving Problem} can be substituted by $\overline{\mathcal{S}}$, called the refined transmit-power-tuple set, such that ${\boldsymbol \mu}_{\max}(\mathbf{s}_t, L, \epsilon) \in \overline{\mathcal{M}}_{[1]}$ holds for all $\mathbf{s}_t \in \overline{\mathcal{S}}$.
Therefore, the branching factor is $|\overline{\mathcal{S}}| = |\overline{\mathcal{M}}_{[1]}|$.
\end{remark}

\textit{Step 2}: More importantly, A* search is employed to further improve the searching efficiency while maintaining the optimality for \probref{prob:Rate-Margin-Deriving Problem}.
A brief description is given here on the application of A* search in solving \probref{prob:Rate-Margin-Deriving Problem}, while we refer the readers to Chapter 3.5.2 in~\cite{RussellS2009BOOK} for a complete description of the A* search algorithm.

For the A* search (or any searching algorithm in general), a node is a fundamental concept.
In our case, the node is $\left(\mathbf{Q}_t, (\mathbf{s}_i)_{i=1}^{t}\right)$, which depends on $\mathbf{Q}_t$ the state, and $(\mathbf{s}_i)_{i=1}^{t}$ is the path to achieve this state from initial node $\left(\mathbf{Q}_0, \emptyset\right)$.
The A* search requires five components to be implemented:
\begin{itemize}
\item   \textbf{Initial node.} The node for starting the search, which is $\left(\mathbf{Q}_0, \emptyset\right)$.
\item   \textbf{Action space.} The set of actions that move from a node to all possible child nodes.
    In our case, the action space is $\overline{\mathcal{S}}$.
\item   \textbf{Goal.} The condition for stopping the search.
    In our case, the goal is $\mathbf{Q}_p = \mathbf{0}$, or simply denoted as $\mathbf{0}$.
\item   \textbf{Step cost.} The step cost is the cost for each searching step.
    In \probref{prob:Rate-Margin-Deriving Problem}, it is
    \begin{align}\label{eqn:Step Cost}
    c_t =
    \begin{cases}
    1 & t < p,\\
    \underset {n\in\mathcal{N}} {\max} \left[\frac {Q_0^{(n)}} {\sum_{t=1}^{p} \mu_{\max}^{(n)}(\gamma_n(\mathbf{s}_t), L, \epsilon)L}\right] & t = p.
    \end{cases}
    \end{align}
    %
\item   \textbf{Evaluation function.} It records the path cost (the summation of step cost) from the past and estimates the path cost in the future.
To be more specific, for a given node $\left(\mathbf{Q}_t, (\mathbf{s}_i)_{i=1}^{t}\right)$, the evaluation function $F(\cdot,\cdot)$ is
\begin{align}\label{eqn:Evaluation Function}
F\left(\mathbf{Q}_t, (\mathbf{s}_i)_{i=1}^{t}\right) = G\left((\mathbf{s}_i)_{i=1}^{t}\right) + E(\mathbf{Q}_t),
\end{align}
where $G\left((\mathbf{s}_i)_{i=1}^{t}\right)$ returns the path cost from initial node to node $\left(\mathbf{Q}_t, (\mathbf{s}_i)_{i=1}^{t}\right)$ and $E(\mathbf{Q}_t)$, called a heuristic function, estimates the path cost from $\left(\mathbf{Q}_t, (\mathbf{s}_i)_{i=1}^{t}\right)$ to the goal $\mathbf{0}$.
The A* search always expands the node with the smallest $F$.
\end{itemize}

It should be noted that the core of the A* search is to construct a function $E(\cdot)$ satisfying $E(\mathbf{Q}_t) \leq E^*(\mathbf{Q}_t)$ for every $\mathbf{Q}_t$, where $E^*(\mathbf{Q}_t)$ is the actual cost from $\mathbf{Q}_t$ to the goal $\mathbf{0}$.
This constructed function is known as the admissible heuristic function in the artificial intelligence literature~\cite{RussellS2009BOOK}.
In this work, we propose the interference-free based heuristic function as follows
\begin{align}\label{eqn:Interference-Free Based Heuristic Function}
E^I\left(\mathbf{Q}_t\right) = \max_{n\in \mathcal{N}} \frac {Q_t^{(n)}} {\mu_{\max}^{(n)}(\gamma'_n(s_{\max}^{(n)}), L, \epsilon) L},
\end{align}
where $t\in\{1,\dotsc,p\}$, $s_{\max}^{(n)} = \max \mathcal{S}^{(n)}$, and
\begin{align}\label{eqn:Interference Free Gamma}
\gamma'_n(s_{\max}^{(n)}) = \frac {h_{nn} s_{\max}^{(n)}} {W_n}, \quad n\in\mathcal{N}.
\end{align}
We call this heuristic function interference-free based, since compared to~\eqref{eqn:SINR}, the expression~\eqref{eqn:Interference Free Gamma} does not consider the interferences from other transmitters.
The following proposition shows that $E^I(\cdot)$ is admissible.

\begin{proposition}\label{prop:Admissibility of Interference-Free Based Heuristic Function for Problem 1}
Let the actual cost to reach the goal $\mathbf{Q}_p = \mathbf{0}$ be
\begin{multline}
E^*\left(\mathbf{Q}_t\right): =\\
\begin{cases}
p - 1 + \underset {n\in\mathcal{N}} {\max} \left[\frac {Q_0^{(n)}} {\sum_{t=1}^{p} \mu_{\max}^{(n)}(\gamma_n(\mathbf{s}_t), L, \epsilon)L}\right] - t & t < p,\\
0 & t = p,
\end{cases}
\end{multline}
where $t \in \{1,\dotsc,p\}$.
Then $E^I\left(\mathbf{Q}_t\right) \leq E^*\left(\mathbf{Q}_t\right)$ holds for every $\mathbf{Q}_t$.
\end{proposition}

\begin{IEEEproof}
See \apxref{apx:Proof of Proposition 5}.
\end{IEEEproof}

\begin{remark}
Based on \propref{prop:Admissibility of Interference-Free Based Heuristic Function for Problem 1}, $E^I(\mathbf{Q}_t)$ in~\eqref{eqn:Interference-Free Based Heuristic Function} provides an A* search for \probref{prob:Rate-Margin-Deriving Problem}, which improves the computational efficiency and maintains the optimality.
Note that the heuristic function in an A* search and the heuristic method in optimization are two totally different concepts: the latter is often suboptimal, while the former is always optimal once it is admissible.
Thus, \propref{prop:Admissibility of Interference-Free Based Heuristic Function for Problem 1} indeed gives the optimality of our search algorithm.
In terms of the computational efficiency, since an A* search reduces the number of nodes to be expanded, it avoids many redundant calculations, which improves the efficiency (see \secref{sec:Simulation Examples}).
We stress that the computational efficiency is high in the cases interferences are strong or zero, because for these cases the optimal choice of node has a smaller $E^I(\mathbf{Q}_t)$ than that in other nodes so that our A* search tends to have significantly fewer steps.
\end{remark}

\textit{Step 3}: Finally, we propose two pruning strategies to further improve the searching efficiency of the A* search:
\begin{itemize}
\item   After a node is selected by the evaluation function~\eqref{eqn:Evaluation Function}, say $(\mathbf{Q}_{t_1},(\mathbf{s}_i)_{i=1}^{t_1})$, we will check whether the condition ``$t_1 = T$ and $\mathbf{Q}_{t_1} \neq \mathbf{0}$'' holds.
    If this condition holds, then we delete this node from the fringe (or called open set, more details can be found in~\cite{RussellS2009BOOK}).
    This is because the condition ``$t_1 = T$ and $\mathbf{Q}_{t_1} \neq \mathbf{0}$'' corresponds to the node whose data queue has not been cleared in the $T$\textsuperscript{th} time slot, and there is no need to expand such a node.
    This consideration is reasonable, since:
    for the rate margin, \algref{alg:Deriving Rate Margin} does not need to know any exact value of $p^*$ for $p^* > T$, i.e., any node with transmission time greater than $T$ is not considered:
    and for the rate-achieving policy deriving, we just need to consider the nodes with transmission time not greater than $T$.
\item   After selecting a node $(\mathbf{Q}_{t_1},(\mathbf{s}_i)_{i=1}^{t_1})$ to expand, we delete those nodes with $t \geq t_1$ but with $({\boldsymbol \mu}_{\max}(\mathbf{s}_i, L, \epsilon))_{i=1}^{t} \preceq ({\boldsymbol \mu}_{\max}(\mathbf{s}_i, L, \epsilon))_{i=1}^{t_1}$ in the fringe, since those nodes' child nodes are suboptimal.
\end{itemize}

To sum up, our algorithm for solving \probref{prob:Rate-Margin-Deriving Problem} is given in~\algref{alg:Solving Problem 1 with A* Search}, where the A* search algorithm, with our pruning strategy, is
\begin{multline}
\mathrm{A}^*(\mathrm{initial\,node},\,\mathrm{action\,space},\,\mathrm{goal},\\\,\mathrm{step\,cost},\,\mathrm{evaluation\,function}).
\end{multline}
We omit the details of the A* search here, since, other than the pruning strategy we already illustrated, the other parts of the A* search algorithm can be found in standard textbooks (e.g.~\cite{RussellS2009BOOK}).

\begin{algorithm}
\begin{footnotesize}
\caption{Solving \probref{prob:Rate-Margin-Deriving Problem} with A* Search}\label{alg:Solving Problem 1 with A* Search}
\begin{algorithmic}[1]
\REQUIRE
    $T$ number of time slots; $N$ the number of transmitter-receiver pairs; ${\boldsymbol \mu}_{[T]}$ the given average rate-tuple;\\
    $\overline{\mathcal{S}}$ the constrained transmit-power-tuple set.
\ENSURE
    $(\mathbf{s}_t^*)_{t=1}^{p^*}$ the optimal solutions for \probref{prob:Rate-Margin-Deriving Problem};\\
    $p^*$ and $\underset {n\in\mathcal{N}} {\max} \left\{{Q_0^{(n)}} / \left[{\sum_{t=1}^{p} \mu_{\max}^{(n)}(\gamma_n(\mathbf{s}_t), L, \epsilon)L}\right]\right\}$ for the optimal objective in \probref{prob:Rate-Margin-Deriving Problem}.
\STATE  $\mathbf{Q}_0 = T {\boldsymbol \mu}_{[T]} L$;
\STATE  $\left[(\mathbf{s}_t^*)_{t=1}^{p^*}, p^*, \underset {n\in\mathcal{N}} {\max} \left\{{Q_0^{(n)}} / {\sum_{t=1}^{p^*} \mu_{\max}^{(n)}(\gamma_n(\mathbf{s}_t^*), L, \epsilon)L}\right\}\right] = \mathrm{A}^*\left(\left(\mathbf{Q}_0, \emptyset\right), \overline{\mathcal{S}}, \mathbf{0}, c_t, F(\cdot)\right)$;
\RETURN $(\mathbf{s}_t^*)_{t=1}^{p^*}$, $p^*$ and $\underset {n\in\mathcal{N}} {\max} \left\{{Q_0^{(n)}} / {\sum_{t=1}^{p^*} \mu_{\max}^{(n)}(\gamma_n(\mathbf{s}_t^*), L, \epsilon)L}\right\}$.
\end{algorithmic}
\end{footnotesize}
\end{algorithm}

\begin{remark}[Measuring the Searching Efficiency]\label{rek:Measure the Searching Efficiency}
We propose the Effective Branching Ratio (EBR) as the metric for evaluating the search efficiency of our solution method of \probref{prob:Rate-Margin-Deriving Problem}:
\begin{align}\label{eqn:EBR}
\mathrm{EBR} = \frac {B} {|\mathcal{S}|},
\end{align}
where $B$ is the effective branching factor of our method, and the $|\mathcal{S}|$ is the branching factor of the original search tree (see the discussion at the beginning of this subsection).
$B$ is a metric on expanded nodes such that if the total number of expanded nodes is $U$, then
\begin{align}
U = \sum_{t=1}^{p^*} B^t.
\end{align}
We can see that $B$ increases with $U$, which means the smaller EBR is, the more efficient in our algorithm performs.
We will use the proposed EBR in \secref{sec:Simulation Examples} to examine the searching efficiency.
\end{remark}

\section{Numerical Results}\label{sec:Simulation Examples}

To corroborate our theoretical results, numerical results are presented.
In this section, firstly, we give two illustrative examples on achievable/unachievable rate-tuples, respectively:
For the achievable rate-tuple, we give the rate-achieving policy and calculate the rate margin followed by the explanation of its meaning.
For the unachievable rate, we calculate the rate margin and explain its meaning.
Secondly, we conduct the Monte Carlo simulation to highlight the computational efficiency of our methods.
Note that in this section, the maximum achievable rate-tuple for transmitter-receiver pair $n \in \mathcal{N}$ is calculated by~\eqref{eqn:Approximation of Maximum Rate under Finite Blocklength Effect}.

Consider the transmission-rate design for a given network with $N = 3$ transmitter-receiver pairs within $T = 5$ time slots, each contains $L = 100$ channel uses.
The following parameters are at hand (the corresponding units are normalized):
The transmit-power sets of these $3$ transmitter-receiver pairs are $\mathcal{S}^{(1)} = \mathcal{S}^{(2)} = \mathcal{S}^{(3)} = \{0, 5\}$, each having an on-off structure.
The power gains are $h_{11} = 0.8$, $h_{22} = 0.7$, $h_{33} = 0.9$, $h_{12} = h_{21} = 0.15$, $h_{13} = h_{31} = 0.25$ and $h_{23} = h_{32} = 0.3$.
The noise powers are $W_1 = W_2 = W_3 = 0.1$.

Now consider whether the rate-tuple ${\boldsymbol \mu}_{[5]} = [0.5, 0.5, 0.5]^{\mathrm{T}}$ can be achieved with error probability $\epsilon = 0.001$.
Using \thmref{thm:Rate-Achieving Policy for All Achievable Rates}, ${\boldsymbol \mu}_{[5]}$ can be achieved, and the rate-achieving policy is $\mathcal{P}_5 = ({\boldsymbol \mu}_t,\mathbf{s}_t)_{t=1}^5$, where ${\boldsymbol \mu}_1 = [0,2.2698,0]^{\mathrm{T}}$, ${\boldsymbol \mu}_2 = [0,0,2.4466]^{\mathrm{T}}$, ${\boldsymbol \mu}_3 = [2.3636,0,0]^{\mathrm{T}}$, ${\boldsymbol \mu}_4 = [0.1364,0.2302,0.0534]^{\mathrm{T}}$, ${\boldsymbol \mu}_5 = \mathbf{0}$, and $\mathbf{s}_1 = [0,5,0]^{\mathrm{T}}$, $\mathbf{s}_2 = [0,0,5]^{\mathrm{T}}$, $\mathbf{s}_3 = [5,0,0]^{\mathrm{T}}$, $\mathbf{s}_4 = [5,5,5]^{\mathrm{T}}$, $\mathbf{s}_5 = \mathbf{0}$.
This result means that the required rate-tuple can be met in this network.
Furthermore, transmission finish in $4$ time slots, since in time slot $5$, all transmitter-receiver pairs transmit nothing.

To maximally utilize the throughput region of this given network, we can linearly scale up ${\boldsymbol \mu}_{[5]}$ so that each transmitter-receiver pair enjoys a rate increase without changing the fairness (i.e., the direction of rate-tuple).
We employ \thmref{thm:Calculation of Rate Margin} (details are shown in \algref{alg:Deriving Rate Margin}) to compute the rate margin $\delta_5({\boldsymbol \mu}_{[5]}) = 1.2554$.
Hence, the boundary rate-tuple is $\delta_5({\boldsymbol \mu}_{[5]}) {\boldsymbol \mu}_{[5]} = [0.6277, 0.6277, 0.6277]^{\mathrm{T}}$.
%
%

Secondly, we consider whether the rate-tuple ${\boldsymbol \mu}'_{[5]} = [0.3, 1, 1]^{\mathrm{T}}$ can be achieved in this network.
Unfortunately, by \thmref{thm:Rate-Achieving Policy for All Achievable Rates}, the rate-tuple ${\boldsymbol \mu}'_{[5]}$ is not achievable, since $\delta_5({\boldsymbol \mu}'_{[5]}) = 0.9079 < 1$.
However, \thmref{thm:Calculation of Rate Margin} says that in the case of not changing the parameters of the network, ${\boldsymbol \mu}'_{[5]}$ should be at least linearly scaled down to $\delta_5({\boldsymbol \mu}'_{[5]}) = 0.9079$ of ${\boldsymbol \mu}'_{[5]}$ to become achievable.
Otherwise, the network should be redesigned (e.g., enlarge the maximum power of transmitter-receiver pairs).

To corroborate the efficiency of our methods, we conduct the Monte Carlo simulations.
The Average Iteration Number (AIN) and the Average Effective Branching Ratio (AEBR) for deriving rate margin are employed to measure the behaviors:
The AIN represents on average how many iterations are required to derive the rate margin (see \thmref{thm:Calculation of Rate Margin}), and the AEBR reveals the searching efficiency for solving \probref{prob:Rate-Margin-Deriving Problem} in every iteration.

The simulation parameters are given as follows.
The number of transmitter-receiver pairs are $N = 3$, and the transmit-power sets are $\mathcal{S}^{(1)} = \mathcal{S}^{(2)} = \mathcal{S}^{(3)} = \{0,1,2\}$.
The power gains are $h_{11} = h_{22} = h_{33} = 0.5$, and $h_{12} = h_{21} = h_{13} = h_{31} = h_{23} = h_{32} = 0.3$.
The noise powers are $W_1 = W_2 = W_3 = 0.1$.
Similar to two above examples, the blocklength is $L = 100$, and the error probability is $\epsilon = 0.001$.
Simulations are conducted for $T \in \{2, 3, 4, 5\}$, and for each $T$, we randomly and uniformly select $1000$ different ${\boldsymbol \mu}_{[T]}$ from $\Lambda_{[\infty]}$ to calculate the rate margin $\delta_T({\boldsymbol \mu}_{[T]})$.
The results are shown in \tabref{tab:Average Iteration Number and the Average Effective Branching Ratio}.

\begin{table}[h]
\caption{Average Iteration Number and the Average Effective Branching Ratio\label{tab:Average Iteration Number and the Average Effective Branching Ratio}}
\centering
\begin{small}
\begin{tabular}{lcccc}
\toprule
~& $T = 2$ & $T = 3$ & $T = 4$ & $T = 5$\\
\midrule
AIN & $1.802$ & $1.844$ & $2.163$ & $2.627$\\
AEBR & $0.385$ & $0.231$ & $0.146$ & $0.095$\\
\bottomrule
\end{tabular}
\end{small}
\end{table}

From \tabref{tab:Average Iteration Number and the Average Effective Branching Ratio}, we can see that the AINs are reasonably small.
For the searching efficiency in each iteration, the AEBRs are small and decrease with $T$.
To give an intuitive illustration, we take $T = 5$ as an example:
AIN equals $2.556$ means that we need $2.556$ iterations on average to derive the rate margin.
AEBR is $0.095$ implies that if we assume $p^* = T = 5$, the total number of nodes (except for the start node) of original search tree is $\sum_{t = 1}^{5} |\mathcal{S}|^t = \sum_{t = 1}^{5} 27^t = 1.490 \times 10^7$, while, for our A* search, only $\sum_{t = 1}^{5} (0.095*27)^t = 180$ number of nodes are expanded on average.
It can be seen that our algorithm significantly improves the computational efficiency.

\section{Conclusion}

In this paper, the finite-horizon throughput region for a wireless multi-user interference network has been studied.
We proposed the rate margin as a metric to determine the achievability of any given rate-tuple and measure the ability to scale up (down) for any achievable (unachievable) rate-tuple so that the resulting rate-tuple is still within (brought back into) the finite-horizon throughput region.
Also, we provided a complete algorithm for finding a rate-achieving policy for any achievable rate-tuple.
Both the rate margin and the rate-achieving policy can be derived very efficiently by using a modified A* search algorithm, where the interference-free based heuristic function plays an important role.
This work represents a significant step towards understanding the network throughput region over a finite time horizon beyond the simplest one-time-slot scenario.
It also demonstrates the fundamental differences in the throughput region between finite and infinite horizon.

More importantly, the presented work serves as the first step to develop more comprehensive results on finite-horizon throughput region in the future:
\begin{itemize}
\item   The rate margin defined in this paper resolves how to do rate-scaling when preserving fairness.
    If some of the transmitter-receiver pairs have more priority for scaling, then a generalization or different definitions of rate margin can be used to reflect the rate scalability from different design perspectives.
    For example, if a transmitter-receiver pair is predominant, then we need to consider the maximum scalability of one component corresponding to this communication pair, while keeping other components unchanged.

\item  It is worth trying to relax or remove the assumption of treating interference as noise.
    If one considers interference decoding (e.g.,~\cite{SatoH1981,RiniS2014}), the finite-horizon throughput region will be enlarged.
    It would be very interesting to consider interference decoding in the finite blocklength regime.

\item   Ultimately, it would be desirable to generalize the finite-horizon throughput region towards an information-theoretic setting which contains all possible coding/decoding strategies.
    One potential approach is the deterministic approximation approach (see~\cite{AvestimehrAS2011,BreslerG2008ETT}) to derive an easy-to-compute approximated finite-horizon throughput region.
\end{itemize}

%

\appendices

\section{Proof of \thmref{thm:Calculation of Rate Margin}}\label{apx:Proof of Theorem 1}

\begin{figure}[h]
\centering
\includegraphics [width=0.95\columnwidth]{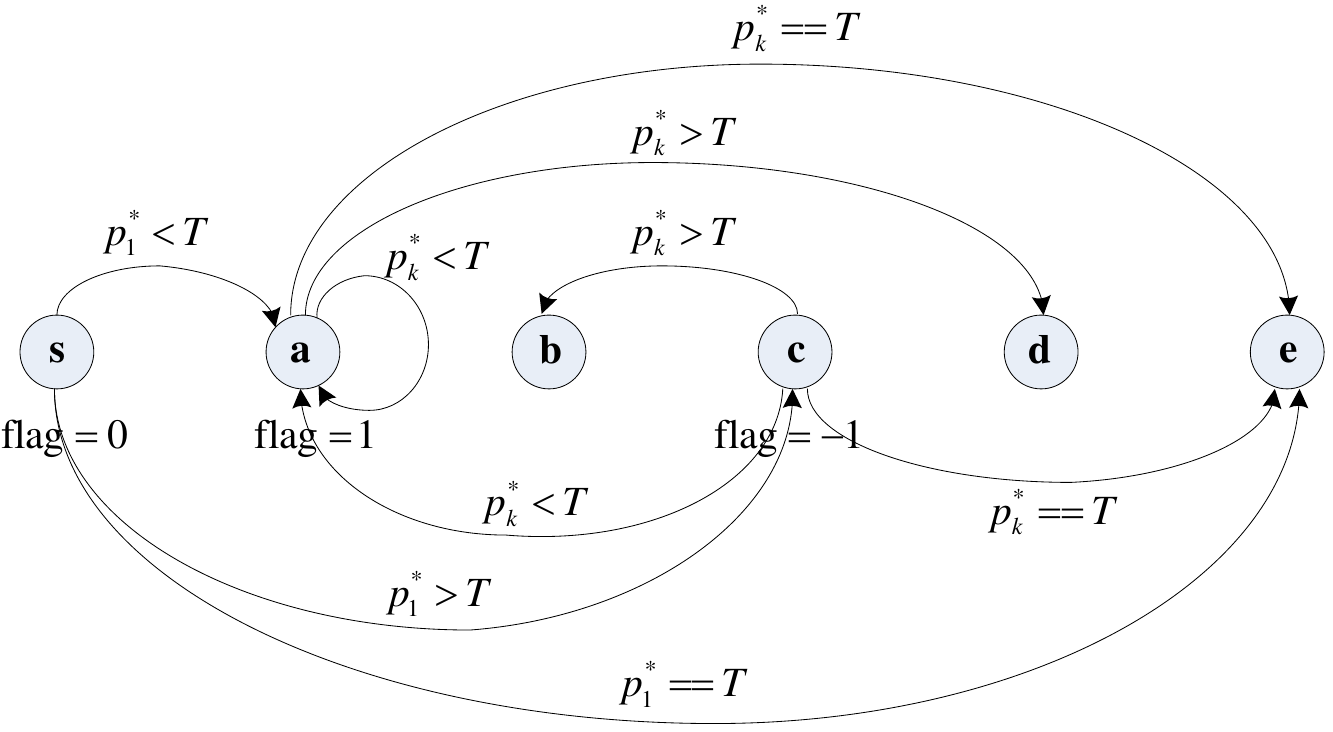}
\caption{A brief version of flow chart for \algref{alg:Deriving Rate Margin}.
The nodes with $\mathbf{a},\mathbf{b},\mathbf{c},\mathbf{d},\mathbf{e}$ represent Lines~\ref{line:a}-\ref{line:end a}, \lineref{line:b}, \lineref{line:c}, \lineref{line:d}, \lineref{line:e} in \algref{alg:Deriving Rate Margin}, respectively, and the node with $\mathbf{s}$ stands for the starting point of \algref{alg:Deriving Rate Margin}.
For $\mathbf{s}$, the value of flag is $0$.
After arriving at node $\mathbf{a}$, the value of flag becomes $1$.
After arriving at node $\mathbf{c}$, the value of flag becomes $-1$.
There are three possible terminals corresponding to nodes $\mathbf{b}$, $\mathbf{d}$, and $\mathbf{e}$.}
\label{fig:Flow chart of Algorithm 1.}
\end{figure}

Before starting the proof, we give a brief flow chart of \algref{alg:Deriving Rate Margin} in \figref{fig:Flow chart of Algorithm 1.}.
With this figure, we can clearly see the flow of \algref{alg:Deriving Rate Margin}: the algorithm starts from $\mathbf{s}$ and ends at three possible terminals $\mathbf{b}$, $\mathbf{d}$, and $\mathbf{e}$ (more details can be found in the caption).
We divide the proof into several cases according to \figref{fig:Flow chart of Algorithm 1.} which is shown as follows.
%

1)~$p^*_1 = T$.
In this case, the program directly goes from node $\mathbf{s}$ to $\mathbf{e}$, and we can easily get $\delta_{T}({\boldsymbol \mu}_{[T]}) = Q_{0,k}^{(1)}/Q_{0,1}^{(1)} = Q_{0,K+1}^{(1)}/Q_{0,1}^{(1)} = \delta_{T}({\boldsymbol \mu}_{[T]})$ in \lineref{line:Deriving Rate Margin}, which means \algref{alg:Deriving Rate Margin} returns the correct result.
Note that $K = 1$.

2)~$p^*_1 < T$.
Initially, the program goes from $\mathbf{s}$ to $\mathbf{a}$.
Then (for $k = 2$), it has three possible destinations, i.e., nodes $\mathbf{a}$, $\mathbf{d}$, and $\mathbf{e}$.
However, the program cannot always stay in node $\mathbf{a}$, and it must end either at $\mathbf{d}$ or $\mathbf{e}$.
This is because $p_k^*$ at least increases by $1$ for each time arriving at $\mathbf{a}$ (recall that $\mathbf{a}$ corresponds to Lines~\ref{line:a}-\ref{line:end a}).
For \lineref{line:a}, if $\lfloor {T}/{p^*_{k}}\rfloor > 1$ or $\rho > 0$, then $\mathbf{Q}_{0,k+1} \succ \mathbf{Q}_{0,k} \delta_{p^*_{k}}({\boldsymbol \mu}_{[p^*_{k}],k})$ (due to ${\boldsymbol \mu}_{[T]} \succ 0$), which implies $\mathbf{Q}_{0,k+1}$ cannot be cleared within $p^*_k$ time slots, and therefore $p^*_{k+1} \geq p^*_k + 1$.
Similarly, if $\lfloor {T}/{p^*_{k}}\rfloor == 1$ and $\rho == 0$, \lineref{line:middle a} returns $\mathbf{Q}_{0,k+1} \succ \mathbf{Q}_{0,k} \delta_{p^*_{k}}({\boldsymbol \mu}_{[p^*_{k}],k})$, and $p^*_{k+1} \geq p^*_k + 1$ still holds.
Hence, the program must stop at node $\mathbf{d}$ or $\mathbf{e}$, and the number of iterations is upper bounded by $K \leq T - p_1^* + 1$, i.e., at least goes to $\mathbf{e}$ (corresponding to $p_K^* = T$).

2-1)~Ends at node $\mathbf{d}$.
From \lineref{line:a}, we know that $p_k^* \leq T$ always holds, because the corresponding $\mathbf{Q}_{0,k+1}$ can always be cleared.
Thus, if the program goes to node $\mathbf{d}$ ($p_k^* > T$), \lineref{line:middle a} must have run, i.e., the increment $\varepsilon T$ makes $p_k^* > T$, where $k = K$.
This means that $\mathbf{Q}_{0,K-1}$ in the $(K-1)$\textsuperscript{th} iteration corresponds to the maximum data queue can be cleared within $T$ time slots.
Then, subtracting the increment $\varepsilon T$ from the current data queue, we can derive $\mathbf{Q}_{0,K+1} = \mathbf{Q}_{0,K-1}$.
Since $\mathbf{Q}_{0,K+1}$ is the maximum data queue that can be cleared within $T$ time slots in the direction of ${\boldsymbol \mu}_{[T]}$, we have $\mathbf{Q}_{0,K+1} = {\boldsymbol \mu}_{[T]} \delta_{T}({\boldsymbol \mu}_{[T]}) T$.
Observe that $\mathbf{Q}_{0,1} = {\boldsymbol \mu}_{[T]} T$, we have $\delta_{T}({\boldsymbol \mu}_{[T]}) = Q_{0,k}^{(1)}/Q_{0,1}^{(1)} = Q_{0,K+1}^{(1)}/Q_{0,1}^{(1)} = \delta_{T}({\boldsymbol \mu}_{[T]})$ in \lineref{line:Deriving Rate Margin}, which means \algref{alg:Deriving Rate Margin} returns the correct result.

2-2)~Ends at node $\mathbf{e}$.
In this case, we directly have $\mathbf{Q}_{0,K+1} = {\boldsymbol \mu}_{[T]} \delta_{T}({\boldsymbol \mu}_{[T]}) T$, and \lineref{line:Deriving Rate Margin} returns the correct rate margin similar to that in 2-1).

3)~$p^*_1 > T$.
Initially, the program goes from $\mathbf{s}$ to $\mathbf{c}$.
Then (for $k = 2$), it has three possible destinations, i.e., nodes $\mathbf{a}$, $\mathbf{b}$, and $\mathbf{e}$.

3-1)~Goes to node $\mathbf{a}$.
This case is similar to 2-1): \algref{alg:Deriving Rate Margin} returns the rate margin correctly, and the iteration number is upper bounded by $K \leq T - p_2^* + 2$ (the program reaches node $\mathbf{a}$ for $k = 2$ rather than $k = 1$).

3-2)~Ends at node $\mathbf{b}$.
In this case, the rate margin $\delta_{T}({\boldsymbol \mu}_{[T]})$ is zero, since in the last iteration $K-1$, the updated data queue is $\mathbf{Q}_{0,K-1+1} = \mathbf{Q}_{0,K} = \varepsilon T$ returned by \lineref{line:c}, and the ``smallest'' rate-tuple $\varepsilon = \mathbf{Q}_{0,K}/T$ (i.e., its component reach the precision of calculation) in the direction of ${\boldsymbol \mu}_{[T]}$ is not achievable.

3-3)~Ends at node $\mathbf{e}$.
This case is similar to 2-2):  \algref{alg:Deriving Rate Margin} returns the rate margin correctly, and the iteration number is upper bounded by $K \leq T - p_2^* + 2$.

\section{Proof of \thmref{thm:Rate-Achieving Policy for All Achievable Rates}}\label{apx:Proof of Theorem 2}

i) $\forall {\boldsymbol \mu}_{[T]} \in \Lambda_{[T]}$, then the data queue can be cleared with some $p \leq T$, which implies $p^* \leq p \leq T$ holds.
Based on $p^* \leq T$, we prove that~\eqref{eqn:Rate-Achieving Policy for All Achievable Rates} is exactly the rate-achieving policy for ${\boldsymbol \mu}_{[T]}$.
By~\eqref{eqn:Rate-Achieving Policy for All Achievable Rates}, the average rate over $T$ slots is
\begin{align}\label{eqn:Average Rate by Rate-Achieving Policy for All Achievable Rates}
\frac {1} {T} \sum_{t=1}^{T} \frac {\mathbf{Q}_{t-1}^* - \mathbf{Q}_{t}^*} {L} = \frac {\mathbf{Q}_{0}} {T L} = \frac {T {\boldsymbol \mu}_{[T]} L} {T L} = {\boldsymbol \mu}_{[T]},
\end{align}
which means the rate is achieved by rate sequence $({(\mathbf{Q}_{t-1}^* - \mathbf{Q}_{t}^*)}/{L})_{t=1}^{p^*}$.
Additionally, since the following holds for every $t \in \{1,\ldots,p^*\}$
\begin{align}
\frac {\mathbf{Q}_{t-1}^* - \mathbf{Q}_{t}^*} {L} \preceq {\boldsymbol \mu}_{\max}(\mathbf{s}_t, L, \epsilon),
\end{align}
the maximum rate constraints (see \defref{def:Rate-Achieving Policy}) are satisfied.
Therefore, ${\boldsymbol \mu}_{[T]}$ can be achieved by the policy $\mathcal{P}_T$.

ii) $\forall {\boldsymbol \mu}_{[T]} \not\in \Lambda_{[T]}$, it follows from \corref{cor:Three Equivalent Statements} that $p^* > T$.

\section{Proof of \propref{prop:Admissibility of Interference-Free Based Heuristic Function for Problem 1}}\label{apx:Proof of Proposition 5}

$\forall \mathbf{Q}_t$, let $\overline{\mathbf{s}}_k = \big[\overline{s}_k^{(1)},\dotsc,\overline{s}_k^{(n)}\big],\, k \in \{t+1,\dotsc,p\}$ be any possible action (transmit power) from $\mathbf{Q}_{k-1}$.
We then divide $E^I\left(\mathbf{Q}_t\right)$ into two parts  to prove the admissibility, i.e., $p - 1 - t$ and $\underset {n\in\mathcal{N}} {\max} \left[{Q_0^{(n)}} / {\sum_{t=1}^{p} \mu_{\max}^{(n)}(\gamma_n(\mathbf{s}_t), L, \epsilon)L}\right]$.
For the first part, $\forall n \in \mathcal{N}$, we have
\begin{align}\label{eqn:Actual Cost to Reach the Goal Part 1}
p - 1 - t = \sum_{k = t + 1}^{p-1} 1 \geq \sum_{k = t + 1}^{p-1} \frac {Q_{k-1}^{(n)} - Q_{k}^{(n)}} {\mu_{\max}^{(n)}(\gamma_n(\overline{\mathbf{s}}_k), L, \epsilon) L}.
\end{align}
Additionally, since $\overline{s}_k^{(n)} \leq s_{\max}^{(n)}$, the following holds
\begin{align}
\gamma_n (\overline{\mathbf{s}}_k) = \frac {h_{nn} \overline{s}_k^{(n)}} {W_n + \sum_{m\neq n} h_{mn} \overline{s}_k^{(m)}} \leq \frac {h_{nn} s_{\max}^{(n)}} {W_n} = \gamma'_n(s_{\max}^{(n)}).
\end{align}
Then, we have $\mu_{\max}^{(n)}(\gamma_n(\overline{\mathbf{s}}_k), L, \epsilon) \leq \mu_{\max}^{(n)}(\gamma'_n(s_{\max}^{(n)}), L, \epsilon)$ for all $n \in \mathcal{N}$.
Thus,~\eqref{eqn:Actual Cost to Reach the Goal Part 1} can be further bounded as
\begin{align}\label{eqn:Actual Cost to Reach the Goal Part 1 with Zoom}
\begin{split}
p - 1 - t &\geq \sum_{k = t + 1}^{p - 1} \frac {Q_{k-1}^{(n)} - Q_{k}^{(n)}} {\mu_{\max}^{(n)}(\gamma_n(\overline{\mathbf{s}}_k), L, \epsilon) L} \\
&\geq \sum_{k = t + 1}^{p - 1} \frac {Q_{k-1}^{(n)} - Q_{k}^{(n)}} {\mu_{\max}^{(n)}(\gamma'_n(s_{\max}^{(n)}), L, \epsilon) L},
\end{split}
\end{align}
for all $n\in\mathcal{N}$, which implies
\begin{align}\label{eqn:Actual Cost Part 1 Zoom Final}
\begin{split}
p - 1 - t &\geq \max_{n\in \mathcal{N}} \sum_{k = t + 1}^{p - 1} \frac {Q_{k-1}^{(n)} - Q_{k}^{(n)}} {\mu_{\max}^{(n)}(\gamma'_n(s_{\max}^{(n)}), L, \epsilon) L} \\
&= \max_{n\in \mathcal{N}} \left[\frac {Q_t^{(n)} - Q_{p-1}^{(n)}} {\mu_{\max}^{(n)}(\gamma'_n(s_{\max}^{(n)}), L, \epsilon) L}\right].
\end{split}
\end{align}
For the second part, we have for all $n \in \mathcal{N}$
\begin{align}\label{eqn:Actual Cost to Reach the Goal Part 2}
\begin{split}
\frac {Q_0^{(n)}} {\sum_{t=1}^{p} \mu_{\max}^{(n)}(\gamma_n(\mathbf{s}_t), L, \epsilon) L} &\stackrel{(a)}{\geq} \frac {Q_{p-1}^{(n)}} {\mu_{\max}^{(n)}(\gamma_n(\mathbf{s}_p), L, \epsilon) L} \\
&\stackrel{(b)}{\geq} \frac {Q_{p-1}^{(n)}} {\mu_{\max}^{(n)}(\gamma'_n(s_{\max}^{(n)}), L, \epsilon) L},
\end{split}
\end{align}
where, for $Q_{p-1}^{(n)} \neq 0$, inequality $(a)$ holds with $Q_0^{(n)} = \sum_{t = 1}^p (Q_{t-1}^{(n)} - Q_t^{(n)})$ and $\sum_{t = 1}^{p-1} (Q_{t-1}^{(n)} - Q_t^{(n)}) = \sum_{t = 1}^{p-1} \mu_{\max}^{(n)}(\gamma_n(\mathbf{s}_t), L, \epsilon) L$.
For $Q_{p-1}^{(n)} = 0$, inequality $(a)$ is satisfied by following that ${Q_0^{(n)}} / {\sum_{t=1}^{p} \mu_{\max}^{(n)}(\gamma_n(\mathbf{s}_t), L, \epsilon)} L$ is nonnegative.
Additionally, inequality $(b)$ holds with $\gamma_{n}\left(\mathbf{s}_{p}\right) \leq \gamma'_n(s_{\max}^{(n)})$.
From~\eqref{eqn:Actual Cost to Reach the Goal Part 2}, we have
\begin{multline}\label{eqn:Actual Cost Part 2 Zoom Final}
\underset {n\in\mathcal{N}} {\max} \left[\frac {Q_0^{(n)}} {\sum_{t=1}^{p} \mu_{\max}^{(n)}(\gamma_n(\mathbf{s}_t), L, \epsilon) L}\right] \\\geq \underset {n\in\mathcal{N}} {\max} \left[\frac {Q_{p-1}^{(n)}} {\mu_{\max}^{(n)}(\gamma'_n(s_{\max}^{(n)}), L, \epsilon)L}\right],
\end{multline}
which, added by~\eqref{eqn:Actual Cost Part 1 Zoom Final}, implies $E^I\left(\mathbf{Q}_t\right) \leq E^*\left(\mathbf{Q}_t\right)$ holds.

\bibliographystyle{IEEEtran}

\bibliography{RateMargin}
\end{document}